\documentclass[aps,twocolumn,superscriptaddress,floatfix]{revtex4}
\usepackage[utf8x]{inputenc}
\usepackage[pdftex]{graphicx}
\usepackage{latexsym}
\usepackage{amsmath}
\usepackage{amsfonts}
\usepackage{amssymb}
\usepackage{bm}
\usepackage{color}
\usepackage{txfonts}
\begin{document}
	\newcommand{\fig}[2]{\includegraphics[width=#1]{#2}}
	\newcommand{\la}{{\langle}}
	\newcommand{\ra}{{\rangle}}
	\newcommand{\dg}{{\dagger}}
	\newcommand{\upa}{{\uparrow}}
	\newcommand{\dna}{{\downarrow}}
	\newcommand{\ab}{{\alpha\beta}}
	\newcommand{\ias}{{i\alpha\sigma}}
	\newcommand{\ibs}{{i\beta\sigma}}
	\newcommand{\hH}{\hat{H}}
	\newcommand{\hn}{\hat{n}}
	\newcommand{\hc}{{\hat{\chi}}}
	\newcommand{\hU}{{\hat{U}}}
	\newcommand{\hV}{{\hat{V}}}
	\newcommand{\br}{{\bf r}}
	\newcommand{\bk}{{{\bf k}}}
	\newcommand{\bq}{{{\bf q}}}
	\def\gsim{~\rlap{$>$}{\lower 1.0ex\hbox{$\sim$}}}
	\setlength{\unitlength}{1mm}
	\newcommand{{\vhf}}{\chi^\text{v}_f}
	\newcommand{{\vhd}}{\chi^\text{v}_d}
	\newcommand{{\vpd}}{\Delta^\text{v}_d}
	\newcommand{{\ved}}{\epsilon^\text{v}_d}
	\newcommand{{\vved}}{\varepsilon^\text{v}_d}
	\newcommand{{\tr}}{{\rm tr}}
	\newcommand{\pprl}{Phys. Rev. Lett. \ }
	\newcommand{\pprb}{Phys. Rev. {B}}

\title { Kagome superconductors AV$_3$Sb$_5$  (A=K, Rb, Cs)}
\author{Kun Jiang}
\affiliation{Beijing National Laboratory for Condensed Matter Physics and Institute of Physics,
	Chinese Academy of Sciences, Beijing 100190, China}

\author{Tao Wu}
\email{wutao@ustc.edu.cn}
\affiliation{Hefei National Laboratory for Physical Sciences at the Microscale, University of Science and
	Technology of China, Hefei, Anhui 230026, China}
\affiliation{CAS Key Laboratory of Strongly-coupled Quantum Matter Physics, Department of Physics, 
	University of Science and Technology of China, Hefei, Anhui 230026, China}

\author{Jia-Xin Yin}
\email{jiaxiny@princeton.edu}
\affiliation{Laboratory for Topological Quantum Matter and Advanced Spectroscopy (B7), Department of Physics, Princeton
	University, Princeton, New Jersey, USA.}

\author{Zhenyu Wang}
\affiliation{Hefei National Laboratory for Physical Sciences at the Microscale, University of Science and
	Technology of China, Hefei, Anhui 230026, China}
\affiliation{CAS Key Laboratory of Strongly-coupled Quantum Matter Physics, Department of Physics, 
	University of Science and Technology of China, Hefei, Anhui 230026, China}

\author{M. Zahid Hasan}
\affiliation{Laboratory for Topological Quantum Matter and Advanced Spectroscopy (B7), Department of Physics, Princeton
	University, Princeton, New Jersey, USA.}

\author{Stephen D. Wilson}
\affiliation{Materials Department and California Nanosystems Institute, University of California Santa Barbara, Santa Barbara,
	California 93106, USA.}

\author{Xianhui Chen}
\email{chenxh@ustc.edu.cn}
\affiliation{Hefei National Laboratory for Physical Sciences at the Microscale, University of Science and
	Technology of China, Hefei, Anhui 230026, China}
\affiliation{CAS Key Laboratory of Strongly-coupled Quantum Matter Physics, Department of Physics, 
	University of Science and Technology of China, Hefei, Anhui 230026, China}

\author{Jiangping Hu}
\email{jphu@iphy.ac.cn}
\affiliation{Beijing National Laboratory for Condensed Matter Physics and Institute of Physics,
	Chinese Academy of Sciences, Beijing 100190, China}
\affiliation{Collaborative Innovation Center of Quantum Matter,
	Beijing 100190, China}
\affiliation{Kavli Institute of Theoretical Sciences, University of Chinese Academy of Sciences,
	Beijing, 100190, China}
\date{\today}

\begin{abstract}
The quasi two-dimensional (quasi-2D) kagome materials AV$_3$Sb$_5$  (A=K, Rb, Cs) were found to be a prime example of kagome superconductors, a new quantum platform to investigate the interplay between electron correlation effects, topology and geometric frustration. In this review, we report recent progress on the experimental and theoretical studies of AV$_3$Sb$_5$ and provide a broad picture of this fast-developing field in order to stimulate an expanded search for unconventional kagome superconductors. We review the electronic properties of AV$_3$Sb$_5$, the experimental measurements of the charge density wave state, evidence of time-reversal symmetry breaking, and other potential hidden symmetry breaking in these materials.  A variety of  theoretical proposals and models that address the nature of the time-reversal symmetry breaking are discussed. Finally, we review the superconducting properties of AV$_3$Sb$_5$, especially the potential pairing symmetries and the interplay between superconductivity and the charge density wave state. 
\end{abstract}
\maketitle

\section{Introduction}
Unveiling new physics from simple lattice models plays a vital role in modern condensed matter physics. For instance, the exact solution of the two-dimensional (2D) Ising model on a square lattice by Onsager revolutionized our view of phase transitions in statistical physics \cite{onsager,huangkerson}; honeycomb lattice graphene can be used to mimic the physics of quantum electrodynamics for Dirac fermions \cite{graphene1,graphene2,geim}. Motivated by the Onsager's solution \cite{onsager}, the Kagome lattice was introduced to statistical physics by Syozi \cite{syozi}, which serves as a rich lattice for realizing novel states and phase behaviors \cite{mekata,yizhou,balents_rev,norman1,norman2}. As shown in Fig.1a, a kagome lattice is formed by corner-sharing triangles.  There are three sublattices labeled as A, B, C, inside each triangle forming the unit cell. Owing to this special lattice structure, the kagome lattice contains geometric frustration for spin systems, which gives rise to extensively degenerate ground states in the nearest neighbor antiferromagnetic Heisenberg model \cite{villain}, as illustrated in Fig.1a. Accordingly, the ground state of the Kagome spin model is the most promising candidate  for the long-sought quantum spin liquid states \cite{yizhou,balents_rev,norman1,norman2,qsl1,qsl2,white,schollwock,jianghc,heyc,xiangtao,sugang,villain}.

Recently, fermionic models on kagome lattices have also become an important platform for studying the interplay among electron-electron correlation effects, band topology and lattice geometry \cite{yin_review}.  The point group of the kagome lattice is the same as graphene \cite{graphene1}, and a standard nearest-neighbor tight-binding (TB) model on the kagome lattice exhibits  Dirac cones at K points, as shown in Fig.1b.   Many distinct properties associated  with Dirac fermions \cite{graphene1} have been discussed, including $\sqrt{n B}$ Landau level \cite{yin20}, tunable Dirac gaps \cite{yin18,ye}, Chern gaps \cite{yin20}, and the quantum anomalous Hall effect \cite{nagaosa,franz} etc.  Besides its Dirac cones, a kagome lattice model can also display  flat bands, as shown in Fig.1b. The flat band arises from the destructive quantum interference of the wave functions from each of the three sublattices. Studying exotic phenomena on flat bands, like fractional Chern insulator states, has been carried out both theoretically and experimentally \cite{flatband1,flatband2,flatband3,fqh,yangbj,cosn1,cosn2}.

\begin{figure}
	\begin{center}
		\includegraphics[width=3.4in]{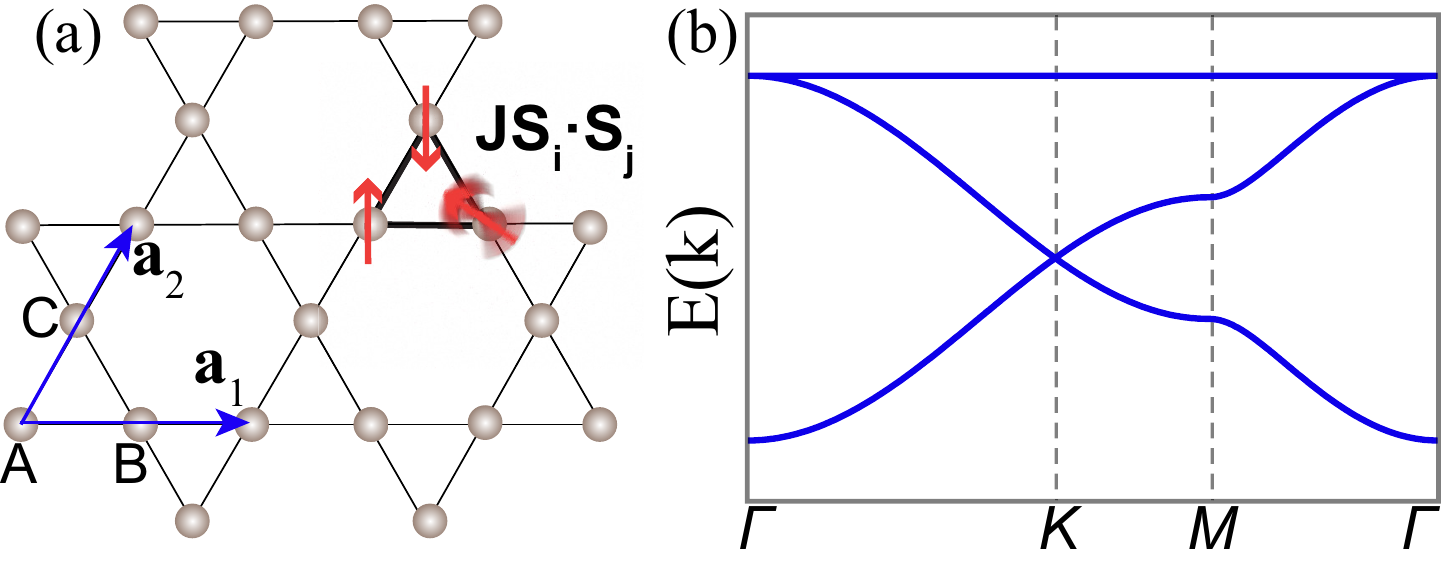}\caption{(a) The crystal structure for the Kagome lattice, which is originated from a Japanese basket-weaving pattern.  The translation vectors are labeled as $\mathbf{a}_1$ and $\mathbf{a}_2$. In each unit cell, there are three sublattices, labeled as A, B, C. For the nearest-neighbor Heisenberg model $\mathbf{J} \mathbf{S}_i \cdot\mathbf{S}_j$, the Kagome lattice faces geometric frustration. As illustrated in the up corner, if two adjacent spins are set to antiparallel, the third spin will face a dilemma. (b) Band structures for nearest-neighbor tight-binding model on Kagome lattice.
			\label{fig1}}
	\end{center}
	\vskip-0.5cm
\end{figure}

In addition to these phenomena,  superconductivity in kagome lattice materials has also been widely discussed. It has been argued that the kagome lattice can host a variety of unconventional pairing superconducting states including $d+id$ chiral superconductor (SC) \cite{jxli,wenxg, qhwang13} and $f$-wave  spin triplet SC \cite{thomale13} among others. However, superconducting kagome materials are rare in nature. Last year, the newly discovered kagome material CsV$_3$Sb$_5$ \cite{ortiz19} was found to be a quasi-2D kagome SC with a transition temperature $T_c \approx 2.3$ K \cite{ortiz20,note}.  Subsequently, superconductivity was also found across the entire family of compounds KV$_3$Sb$_5$ ($T_c \approx 0.93$K) \cite{ortiz21}, and RbV$_3$Sb$_5$ ($T_c \approx 0.75$K) \cite{lei}. This discovery has stimulated extensive research activity in this field \cite{ortiz19,ortiz20,ortiz21,lei,shiyanli,hall20,edgecurrent,graf,topocdw,xhchen}.  

In this review, we  discuss the recent progress in studying this newly discovered AV$_3$Sb$_5$ kagome family. This paper is organized as follows:
We  first discuss the crystal structure and the electronic properties of AV$_3$Sb$_5$ (A=K, Rb, Cs).  Second, we review both the experimental evidence and theoretical understanding of the unconventional  charge density wave order that forms and reports of accompanying time reversal symmetry breaking.  Third, we report the current status of understanding the SC properties of  AV$_3$Sb$_5$.  Finally, we address other unconventional features in these compounds, such as pairing density wave order, and provide future research perspectives.

\section{Crystal and Electronic structures}
\begin{figure*}
	\begin{center}
		\includegraphics[width=7.0in]{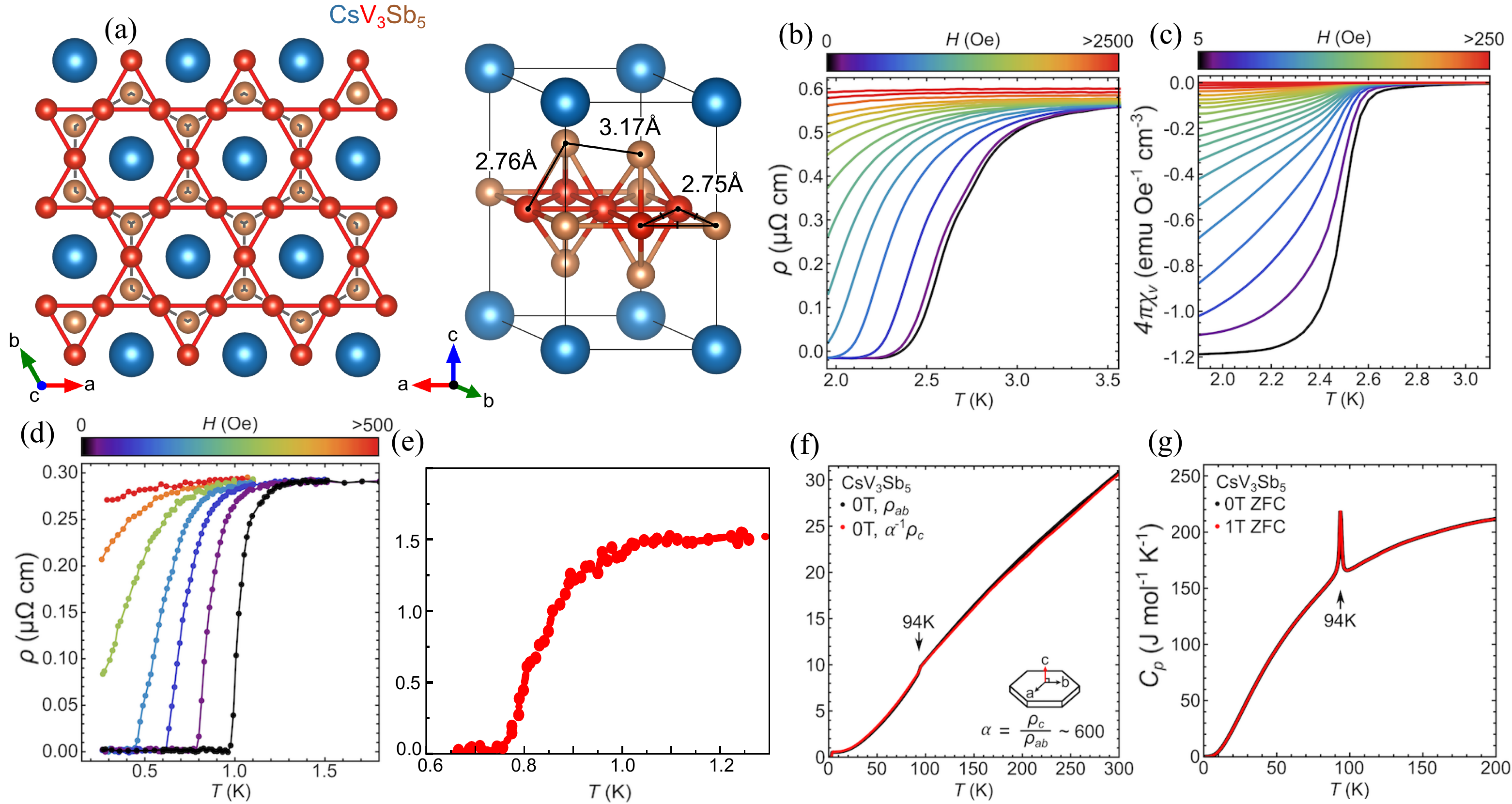}\caption{(a) The crystal structure for CsV$_3$Sb$_5$ \cite{ortiz20}.  (b) (c) Field-dependent resistivity and magnetization at low temperatures, showing the onset of superconductivity  for CsV$_3$Sb$_5$ with $T_c \approx$ 2.3K \cite{ortiz20}. (d) Field-dependent resistivity  at low temperatures for KV$_3$Sb$_5$ \cite{ortiz21}. (e) Resistivity at low temperatures for RbV$_3$Sb$_5$ \cite{lei}. (f) (g) The temperature dependent electrical resistivity, and heat capacity at higher temperature for CsV$_3$Sb$_5$ showing a transition around 94K \cite{ortiz20}.
			\label{fig2}}
	\end{center}
	\vskip-0.5cm
\end{figure*}

The AV$_3$Sb$_5$ materials crystallize into the $P6/mmm$ space group and exhibit a layered structure of V-Sb sheets intercalated by K/Rb/Cs, as shown in Fig.\ref{fig2}(a,b) \cite{ortiz19}.
In the V-Sb plane, three V atoms form the kagome lattice and an additional Sb atom forms a triangle lattice located at the V kagome lattice's hexagonal center. This V kagome layer largely dominates the physics behind AV$_3$Sb$_5$ as discussed later. Above and below the V-Sb plane, out-of-plane Sb atoms form two honeycomb lattice planes respectively with lattice sites located above and below the centers of the V triangles in the kagome plane. A-site atoms form another triangular lattice above or below these Sb honeycomb or antimonene planes.

We can first understand the electronic properties of AV$_3$Sb$_5$ from the transport measurements. The low temperature electrical resistivity $\rho (T)$ and its field-dependence  are plotted in Fig.\ref{fig2}(b) for CsV$_3$Sb$_5$ \cite{ortiz20}. One finds that the zero-field $\rho (T)$  shows a broad transition towards the SC ground state with $T_c \approx 2.3$K, which is continuously suppressed by applying a magnetic field. The magnetization data in Fig.\ref{fig2}(e) also reveals a well-defined Meissner effect, and heat capacity measurements show a sharp entropy anomaly at the SC transition \cite{ortiz20}. Therefore, the CsV$_3$Sb$_5$ becomes the first example of quasi-2D kagome SCs.  The critical field $H_c$ for CsV$_3$Sb$_5$ is relatively small with the c-direction $H_{c2}\approx$ 0.4T \cite{wenhaihu,dongxiaoli}. Similarly, the $\rho (T)$ of KV$_3$Sb$_5$ drops to zero with $T_c \approx 0.93$K shown in Fig.\ref{fig2}(e) \cite{ortiz21} and RbV$_3$Sb$_5$ has a $T_c \approx 0.75$K \cite{lei}. Hence, all AV$_3$Sb$_5$ compounds within the material family are superconducting at low temperature.
 
Above the SC ground state, the normal states of AV$_3$Sb$_5$ also show quite different behavior. The temperature-dependent resistivity of KV$_3$Sb$_5$ can be modeled by a Fermi-liquid formula $\rho (T)=\rho_0+aT^2$ \cite{ortiz19}, which shows a typical metallic behavior. The in-plane and out-plane resistivity data show a large anisotropy with a ratio $\alpha=\frac{\rho_c}{\rho_{ab}}\approx 600$ in CsV$_3$Sb$_5$, as shown in Fig.\ref{fig2}(f) \cite{ortiz20}. This large anisotropy agrees well with the quasi-2D nature of AV$_3$Sb$_5$, where the V kagome layers play a dominant role in the electronic properties. Hence, the AV$_3$Sb$_5$ is a quasi-2D metal.  $\rho(T)$ also contains a kink behavior around $94$ K, which is  related to the long-range CDW order discussed later \cite{ortiz20}. A sharp peak from the heat capacity data at this same temperature indicates the CDW transition is a first-order phase transition \cite{ortiz20}, where the first derivatives of free energy are not continuous.  The lack of phonon softening near this transition from the inelastic x-ray scattering  also suggests that the transition is weakly first order \cite{miaohu_xrd,miaohu2}. It is worth mentioning that this weakly first order transition is best characterized in CsV$_3$Sb$_5$, and the nature of the transition merits further study in the other compounds.

\begin{figure*}
	\begin{center}
		\includegraphics[width=7.0in]{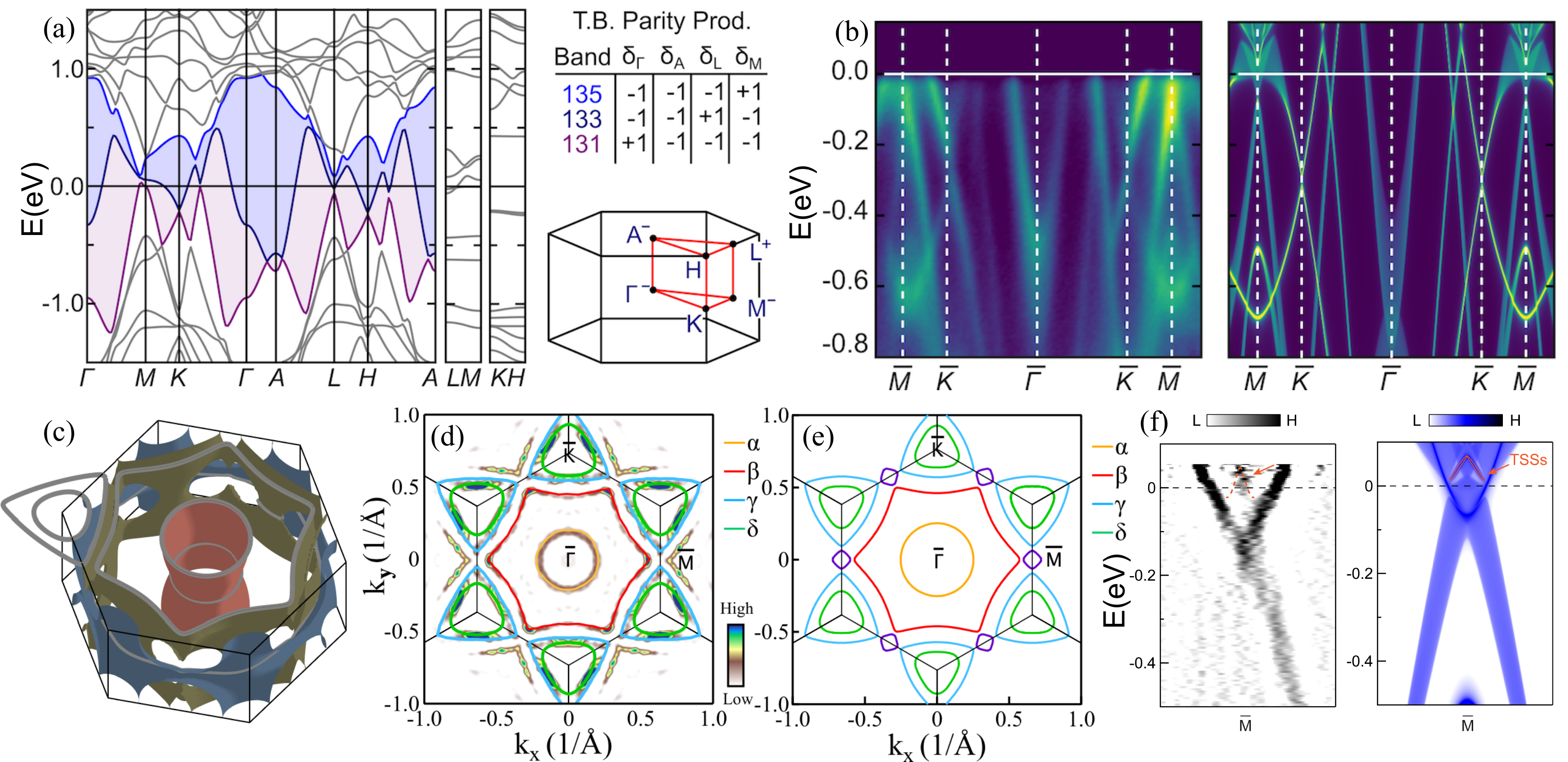}\caption{(a) The band structure of CsV$_3$Sb$_5$ calculated by DFT. The insert shows the parity eigenvalues for each band at the time-reversal invariant momentum points \cite{ortiz20}. (b) ARPES measured band structure (left) and its comparation with DFT (right) for CsV$_3$Sb$_5$ \cite{ortiz20}. (c) FS calculated for CsV$_3$Sb$_5$ at experimetal $E_F$ \cite{wilson_quantum}. (d) (e) FSs measured by ARPES and calculated by DFT for KV$_3$Sb$_5$ \cite{zhouxingjiang}. (f) The ARPES measured (left) and DFT calculated (right) topological surface states (TSSs)  for CsV$_3$Sb$_5$ \cite{shiming}.
			\label{fig3}}
	\end{center}
	\vskip-0.5cm
\end{figure*}

To reveal the electronic nature of AV$_3$Sb$_5$, density functional (DFT) calculations and angle-resolved photoemission spectroscopy (ARPES) measurements have been performed \cite{ortiz20,ortiz19,borisenko,comin,hejunfeng,miaohu_xrd,shendawei,shiming,wangshancai,zhangyan,zhouxingjiang,wilson_quantum,sato}.
The DFT calculations show multiple bands crossing the Fermi level ($E_F$) in CsV$_3$Sb$_5$, as shown in Fig.\ref{fig3}(a). Around the $\Gamma$ point, there is an electron-like parabolic band, which originates from the in-plane Sb $p_z$ orbital. The bands around the Brillouin zone (BZ) boundaries are mainly attributed to the V $d$ orbitals.  Notice that, there are two van Hove (VH) points close to $E_F$ around the M point, which play an important role in the symmetry breaking observed in AV$_3$Sb$_5$.  The upper VH  point is further connected with the Dirac cone around the K point, which reflects a typical feature of the kagome model described above. ARPES measurements find that  the electronic band structure of CsV$_3$Sb$_5$ qualitatively agrees with DFT calculations \cite{ortiz20}, as compared in Fig.\ref{fig3}(b), and DFT calculations provide qualitatively accurate descriptions of the electronic structures of AV$_3$Sb$_5$ systems.

To confirm the quasi-2D nature of AV$_3$Sb$_5$, the three dimensional Fermi surface (FSs) of CsV$_3$Sb$_5$ is calculated in Fig.\ref{fig3}(c). The FSs show the traditional cylinder behaviors as in copper-based and iron-based superconductors  \cite{hussey,iron_book,keimer}, which is the origin of large resistivity anisotropy. The excellent agreement between DFT and ARPES  indicates a small band renormalization owing to correlation effects in the lattice. Hence, the AV$_3$Sb$_5$ materials are effectively modeled as weakly correlated systems \cite{wangyl}. For example, the high-resolution ARPES data from KV$_3$Sb$_5$ find excellent matching between the measured and calculated FSs \cite{zhouxingjiang}, as plotted in Fig.\ref{fig3}(d,e).
 
 Besides the above electronic structures, CsV$_3$Sb$_5$ also carries a non-trivial Z$_2$ topological index \cite{ortiz20}. For inversion symmetric and time-reversal symmetric systems,  the Z$_2$ topological invariant can be obtained from time-reversal invariant momentum points with their inversion operator eigenvalues \cite{z2}. As listed in Fig.\ref{fig3}(a), the Z$_2$ invariant is nontrivial for band numbers 131,133, 135 enumerated in DFT calculations. The parity index for 133, 135 bands at $M$ point is different, which gives rise to a band inversion at $M$. Therefore, the normal state of CsV$_3$Sb$_5$ is a Z$_2$ topological metal, and this $Z_2$ topological property leads to a surface state embedding around the bulk FS at the M point. ARPES experiments have resolved this feature as shown in Fig.\ref{fig3}(f).

\section{Charge Density Wave and Symmetry Breaking}
\begin{figure*}
	\begin{center}
		\includegraphics[width=7.0in]{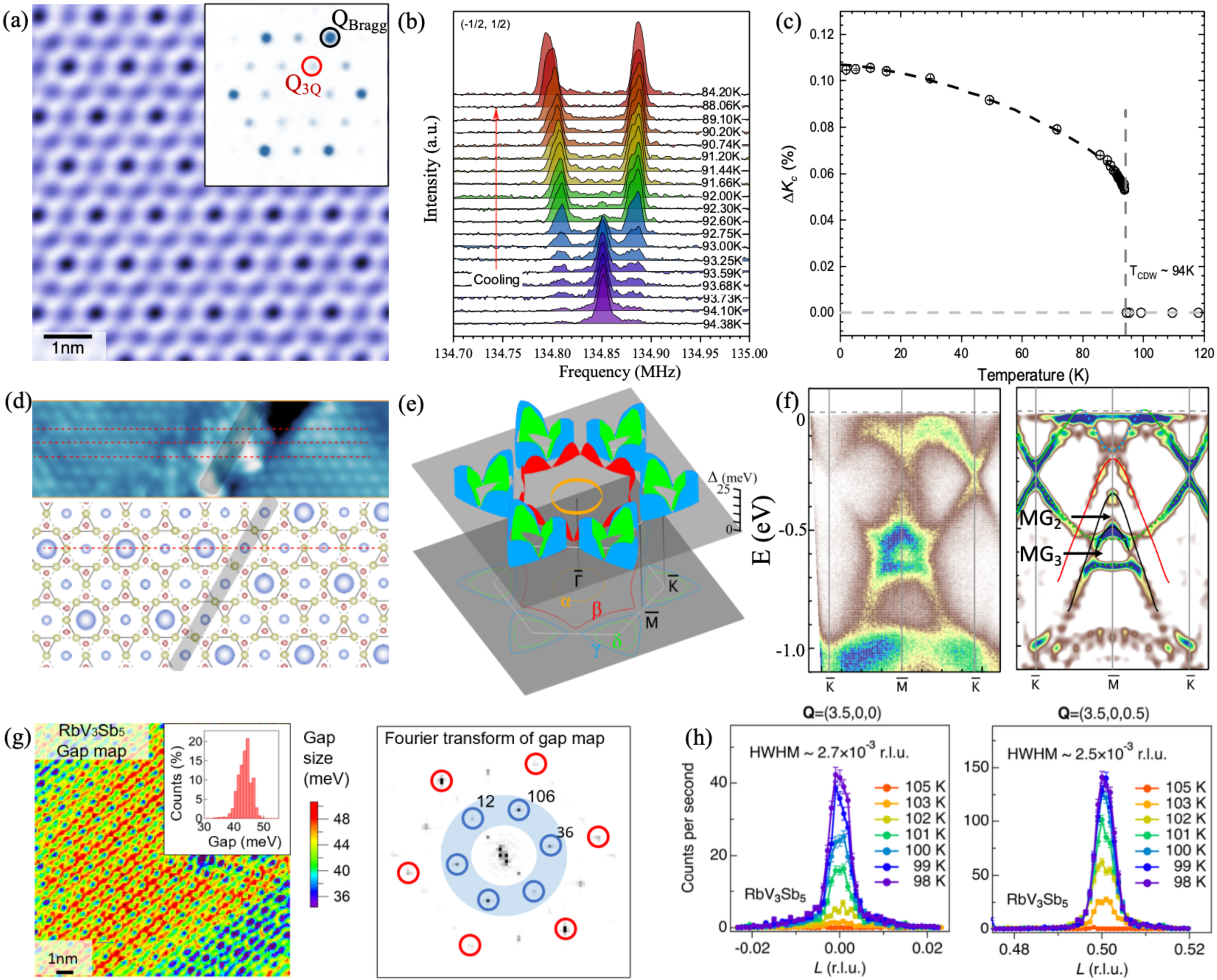}\caption{(a) A topographic image of a large Sb surface and its Fournier transformation showing a 2$\times$2 modulation for KV$_3$Sb$_5$ from STM. Besides the Bragg peaks $Q_{Bragg}$, there are additional charge modulation peaks $Q_{3Q}$ \cite{topocdw}. (b) The temperature dependence of  the central transition lines of  $^{51}$ V NMR  with the temperature cooling across $T_{CDW}$ for CsV$_3$Sb$_5$ \cite{wutao}. (c) Temperature dependence of the splitting of Knight shift $\Delta K_c$  for CsV$_3$Sb$_5$ \cite{wutao}.
			(d) The STM scanning of the step edge in  CsV$_3$Sb$_5$. The dashed pink lines track the chains with CDW modulation on the upper side. A $\pi$-phase jump can be observed between the upper and lower sides. The illustration of the CDW patterns near a single-unit-cell step is plotted in the lower panel \cite{zywang}.  (e) The CDW gap structures for each FSs in KV$_3$Sb$_5$ measured by ARPES \cite{zhouxingjiang}. (f) ARPES measured band structures (right) and their  second derivatives along $\bar{K}-\bar{M}-\bar{K}$. There are two additional CDW gaps at $MG_2$ and $MG_3$  away from $E_F$ \cite{zhouxingjiang}. (g) Real-space CDW gap map for RbV3Sb5 and its Fourier transform. The 2$\times$2 vector peaks show different intensities, defining a kind of electronic chirality \cite{yaoyugui}.
			(h) The temperature dependent CDW peaks of RbV$_3$Sb$_5$ at $Q = (3.5, 0, 0)$ and $(3.5, 0, 0.5)$. The CDW peak at half-integer L demonstrates a 3D CDW with 2 $\times$2$\times$  2 superstructure \cite{miaohu_xrd}.
			\label{fig4}}
	\end{center}
	\vskip-0.5cm
\end{figure*}

As discussed in the previous section, a CDW phase transition occurs for all AV$_3$Sb$_5$ materials ranging from $78$ to $103$K ($T_{CDW}\approx$94K for CsV$_3$Sb$_5$, $T_{CDW}\approx$103K for RbV$_3$Sb$_5$, $T_{CDW}\approx$78K for KV$_3$Sb$_5$) \cite{ortiz19,ortiz20,ortiz21,lei}. In the first report of AV$_3$Sb$_5$ crystal growth, elastic neutron scattering measurements ruled out the possibility of long-range magnetic order \cite{ortiz19}. The absence of long-range magnetic order was further confirmed by the muon spin spectroscopy, indicating the transition derives primarily from the charge degree of freedom \cite{graf}. Soon after SC was discovered in CsV$_3$Sb$_5$, scanning tunneling microscopy (STM) measurements were performed on the Sb and K surfaces of KV$_3$Sb$_5$, revealing that the transition is a CDW transition with 2 $\times$ 2 superlattice modulation \cite{topocdw,fengdonglai,hezhao,zywang,lihong,gaohongjun,shumiya,yaoyugui}. From the STM topographic spectrum in Fig.\ref{fig4}(a), the charge modulation on the Sb surface is resolved \cite{topocdw}. By Fourier transforming the topographic image, there are six additional ordering peaks Q$_{3Q}$ in addition to those from the primary lattice structure \cite{topocdw}.
STM further shows an energy gap opened around the Fermi energy of  $\sim$50meV, which together with the 2 $\times$ 2 superlattice modulation disappears above T$_{CDW}$ \cite{topocdw,fengdonglai,hezhao,zywang,lihong,gaohongjun,shumiya,yaoyugui}. Across this gap, there is a real-space charge reversal for the 2 $\times$ 2 superlattice modulation \cite{topocdw}, which is a hallmark of CDW ordering.
 
 Nuclear magnetic resonance (NMR) measurements further support the absence of magnetic order and confirm that the CDW transition is indeed a first order transition \cite{wutao}. From the NMR spectrum, there  are two V signals after the CDW transition, V(I) and V(II) as shown in the insert of Fig.\ref{fig4}(b). The splitting of Knight shift $\Delta K_c$ between V(I) and V(II) sites shows a sudden jump at $T_{CDW}$. Beyond the surface sensitive measurements, the CDW state is found to be three-dimensional and be modulated along the $c$-axis. 
This modulation is either 2 $\times$ 2  $\times$ 2 or 2 $\times$ 2  $\times$ 4  for AV$_3$Sb$_5$ materials with 2 $\times$ 2  $\times$ 2 reported for KV$_3$Sb$_5$ and both 2 $\times$ 2  $\times$ 2 and 2 $\times$ 2  $\times$ 4 reported for CsV$_3$Sb$_5$ \cite{miaohu_xrd,topocdw,wilson_quantum,miaohu2}, as shown in Fig.\ref{fig4}(g,h). Disorder along the $c$-axis impacts crystallinity in the direction of the out-of-plane modulation and potentially accounts for this discrepancy.  The 3D modulation is also confirmed by the STM data collected across surface step edges \cite{zywang} and a $^{133}Cs$ NMR spectrum study \cite{wutao}.   Future studies are underway to fully understand the $c$-axis periodicity of the superlattice. On the clean surface regions of CsV$_3$Sb$_5$ and RbV$_3$Sb$_5$, STM detects real-space modulations of the CDW gap as shown in Fig.\ref{fig4}(g). Interestingly, the Fourier transform of the gap map also shows the 2 $\times$ 2 vector peaks with different intensities, thus revealing a novel electronic chirality of the CDW order \cite{yaoyugui,topocdw}.


In order to determine the gap structures in  momentum space, several high-resolution ARPES measurements have been performed \cite{sato,zhouxingjiang,wangshancai,zhangyan,miaohu_xrd}. Based on ARPES data, we can find that different FSs in KV$_3$Sb$_5$ exhibit diverse CDW gap structures, as shown in Fig.\ref{fig4}(e). The CDW gap vanishes for the $\alpha$ FS around the BZ $\Gamma$ point. Since the $\alpha$ FS  stems from the $p_z$ band of the in-plane Sb, the $p_z$ orbital does not participate in the CDW formation \cite{zhouxingjiang,zhangyan}.  In contrast, the V-derived FSs around the BZ boundary exhibit highly momentum-dependent CDW gaps, which are dominated by quasiparticles around the van Hove singularities at the M points \cite{zhouxingjiang,zhangyan}.  Quantum oscillation measurements also support the dominant role of vanadium orbitals within the CDW order \cite{wilson_quantum}. Hence, the V kagome layer dominates the CDW gaps and the VH quasiparticles deeply influence the gap structure in AV$_3$Sb$_5$.
In addition to the gaps resolved around the FSs, ARPES data in KV$_3$Sb$_5$ have also observed large CDW gap opening below $E_F$ \cite{zhouxingjiang}. For instance, at the M point, a 150 meV gap opens at the MG$_2$  and a 125 meV gap for MG$_3$ at 20 K, as shown in Fig.\ref{fig4}(f). This feature strongly indicates that the structural transition plays an important role in this CDW transition.  It is also clear that the structural transition mostly affects the V kagome network, while the out-of-plane coupling involving Sb $p_z$ orbitals is hardly changed. 

\subsection{Time-reversal symmetry breaking}
\begin{figure*}
	\begin{center}
		\includegraphics[width=7.0in]{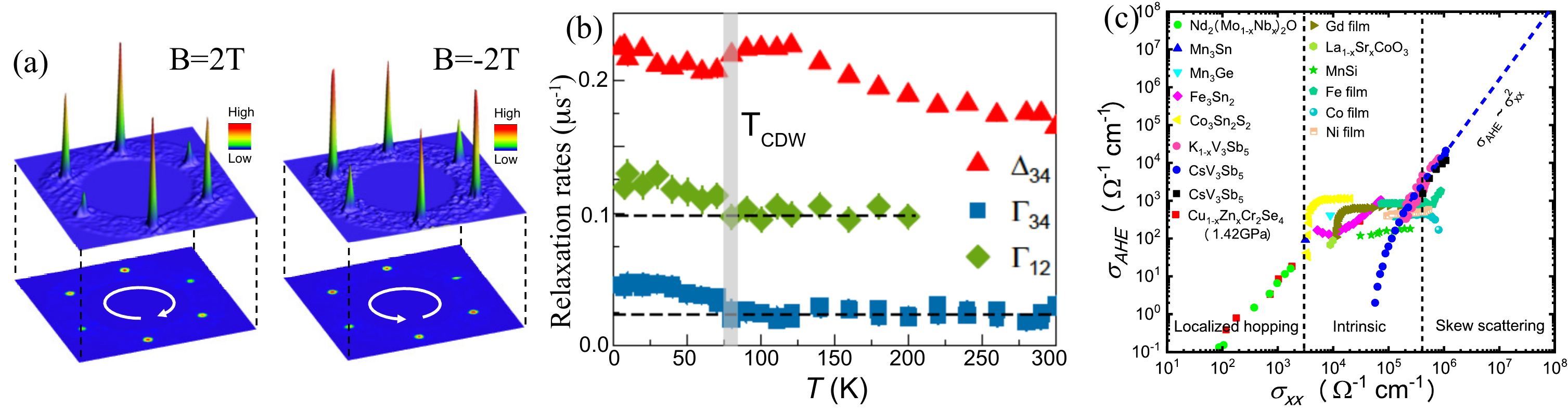}\caption{(a) Spectroscopic 2$\times$2 vector peaks for KV$_3$Sb$_5$ taken at B=2T and B=-2T, respectively. The highest vector peaks shift their positions under magnetic field \cite{topocdw}.
			 (b) The temperature-dependent  muon relaxation rates in KV$_3$Sb$_5$. The $\Gamma_{12}$ measures the rates collected in the forward and backward detectors while the $\Gamma_{34}$ and $\Delta_{12}$ measure the rates collected in the up and down detectors.  The relaxation rates start to increase below the CDW transition \cite{musr1}.  (c) $\sigma_{AHE}$ versus $\sigma_{xx}$ for a variety of materials compared with CsV$_3$Sb$_5$ spanning various regimes from localized hopping regime to the skew scattering regime \cite{xhchen}. 
			\label{fig5}}
	\end{center}
	\vskip-0.5cm
\end{figure*}

Interestingly, accumulated evidence for  time-reversal symmetry breaking (TRSB) signals was found in the CDW phases of AV$_3$Sb$_5$ compounds. Since charge is a  time-reversal symmetry preserving quantity, the emergence of this TRSB becomes one of the more intriguing phenomena in these otherwise non-magnetic AV$_3$Sb$_5$ materials.
The first evidence for TRSB was found in a magnetic field dependent STM measurements \cite{topocdw}. As discussed above, there are six CDW ordering vectors  Q$_{3Q}$ from the STM topographic spectrum. 
However, the intensity of these three pairs of vectors are different in the clean regions for all AV$_3$Sb$_5$ materials \cite{topocdw,hezhao,yaoyugui}, thus defining a chirality of the CDW order (counting direction from the lowest intensity peak pairs to highest intensity peak pairs). The chirality of the CDW order further shows an unusual response to the perturbation of external magnetic field B. As shown in Fig. \ref{fig5}(a), the chirality switch from anticlockwise to clockwise when the magnetic fields changes from +2T to -2T applied along the c-axis. 
Owing to the Onsager reciprocal relation, the response functions of a time-reversal preserving system under $+B$ and $-B$ must relate to each other by a time-reversal operator. This non-reciprocal relation under magnetic field breaks the Onsager relation indicating the TRSB in this non-magnetic kagome system \cite{topocdw}.

The strongest evidence for TRSB comes from the zero-field muon spin relaxation/rotation ($\mu$SR) spectroscopy \cite{musr1,yuli}. The spin-polarized muons were implanted into the AV$_3$Sb$_5$ single crystals. The muon spin will rotate and relax under the influence of local magnetic fields. The $\mu$SR technique is highly sensitive to the extremely small magnetic fields, capable of detecting on the order of 0.1 Gauss fields experienced by the implanted muons. 
As shown in Fig.\ref{fig5}(b), the relaxation rates of KV$_3$Sb$_5$ start to increase below the CDW transition temperature $T_{CDW}$, which strongly suggests the emergence of a local magnetic field owing to TRSB \cite{musr1}.  
Similar measurements on CsV$_3$Sb$_5$ also found TRSB signals \cite{yuli}. However, the TRSB transition temperature is slightly lower than the  $T_{CDW}\approx90$K. We will come back to discuss the physical origin of this TRSB  in the next section.

Moreover, a giant anomalous Hall effect (AHE) has also been observed in AV$_3$Sb$_5$ \cite{hall20,xhchen}, and the onset of this AHE was found to be concurrent with the CDW order \cite{xhchen}.
Normally, there are two origins of the AHE, intrinsic Berry curvature and extrinsic impurity scattering \cite{nagaosa_ahe}. As shown in Fig.\ref{fig5}(c), by comparing transverse $\sigma_{AHE}$ and longitudinal $\sigma_{xx}$ conductivity, both the intrinsic Berry curvature  and the impurity-induced skew scattering contribute to the giant AHE in KV$_3$Sb$_5$ and CsV$_3$Sb$_5$. However, compared to conventional spontaneous AHE with  ferro- or ferrimagnetic ordering, the AHE in  AV$_3$Sb$_5$ exhibits $\sigma_{AHE}(B \rightarrow 0)= 0$ without a hysteresis behavior. The $\sigma_{AHE}(B \rightarrow 0)=0$ feature might originate from the anti-phase TRSB between adjoining kagome layers or domain walls \cite{yuli}.  Besides the above $\mu$SR, AHE, and magnetic field-dependent STM measurements,  results from other TRSB sensitive techniques like the polarized neutron diffraction and Kerr effect etc. are highly desired.

\subsection{Inversion Symmetry and Nematicity}
Besides the translation symmetry breaking and time-reversal symmetry breaking associated with the CDW state, what are the remaining symmetries within the CDW state becomes an interesting question. The point group of  AV$_3$Sb$_5$ $P6/mmm$ space group is $D_{6h}$, which can be generated by the $C_6$ rotation, inversion operator ${\cal I}$   and the mirror operator  $\sigma_x$ about the $yz$ plane \cite{fengxl2}. 

To test the inversion symmetry ${\cal I}$, second-harmonic generation (SHG) optical data were collected from CsV$_3$Sb$_5$ \cite{yuli}. SHG measures the second-order nonlinear optical response $ \mathbf{P}= \epsilon_0\chi^{(2)} \mathbf{E} \mathbf{E}$, where $\mathbf{P}$ is the electric polarization induced by the incident light with the electric field $\mathbf{E}$ and $\epsilon_0$ is the vacuum permittivity. Since $\mathbf{P}$  and $\mathbf{E}$ are odd under inversion symmetry ${\cal I}$, the rank-three nonlinear optical susceptibility tensor $\chi^{(2)}$ is only finite when parity is broken. Only negligibly small SHG signals (likely originating from the surface) were detected from 120 K down to 6K. Hence, inversion symmetry ${\cal I}$ remains a valid symmetry for AV$_3$Sb$_5$ at all temperatures, which constrains the CDW order and will be also important for superconducting pairing possibilities discussed in the following section. 

\begin{figure*}
	\begin{center}
		\includegraphics[width=7.0in]{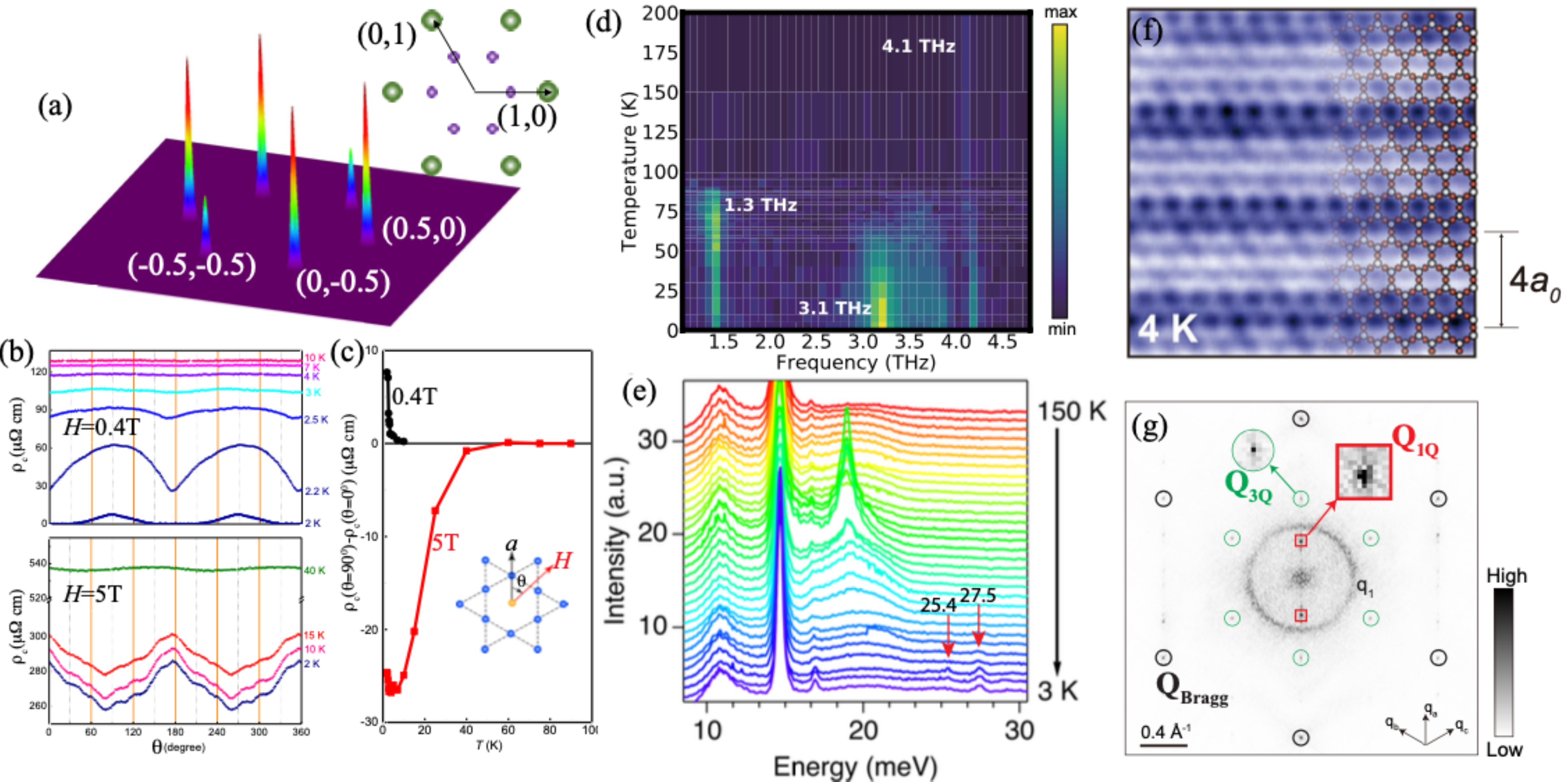}\caption{(a) Spectroscopic 2$\times$2 vector peaks for KV$_3$Sb$_5$ taken at zero external field \cite{topocdw,miaohu2}.  (b) Angular dependent c-axis resistivity for CsV$_3$Sb$_5$ measured at different temperatures under a magnetic field of  0.4 T (up panel) and 5 T (down panel) \cite{wenhaihu}. (c) Temperature dependence of nematicity of c-axis resistivity between $\theta=0^{\circ}$ and $90^{\circ}$ \cite{wenhaihu}. (d) Temperature dependence waterfall map of  coherent phonon spectroscopy for CsV$_3$Sb$_5$ \cite{wangnanlin}.  The 4.1 THz coherent phonon is present at all temperatures through phase change. The 1.3 THz phonon can be only detected below T$_CDW$, while the 3.1 THz phonon only shows up at the temperature below $30\sim60$K. (e) Raman spectroscopy for KV$_3$Sb$_5$. Below 30 K, two new phonon modes at 25.4 and 27.5 meV were observed \cite{miaohu_xrd}. (f) (g) 1$\times$ 4 charge modulation and its Fournier transformation found in the Sb surfaces of CsV$_3$Sb$_5$. In (g), there are two $Q_{1Q}$ peaks in addition to Q$_{Bragg}$ and Q$_{3Q}$ \cite{hezhao}.
			\label{fig6}}
	\end{center}
	\vskip-0.5cm
\end{figure*}

Rotational symmetry breaking without translational symmetry breaking, namely the nematicity, is another important issue for understanding unconventional electron liquids \cite{nematic1,nematic2}. 
For KV$_3$Sb$_5$, low temperature STM data above SC $T_c$ at zero-field found that the CDW peak intensities at Q$_{3Q}$ show a $C_6$ rotation broken feature \cite{topocdw,lihong,miaohu2}, as shown in Fig.\ref{fig6}(a) as a simulation of $2\times2$ vector peaks on the surface based on bulk $2\times2\times2$ CDW.  
Magnetoresistance measurements in CsV$_3$Sb$_5$ also reveal  the nematic nature of the CDW state persisting into the superconducting phase \cite{wenhaihu,dongxiaoli}, as shown in Fig.\ref{fig6}(b).
Therefore, the CDW state is electronically nematic with only $C_2$ rotation symmetry at low-temperature. 
 Notice that the z-direction modulated CDW reduces the point group symmetry from $D_{6h}$ down to $D_{2h}$ \cite{miaohu2,fengxl2}. However, from the magnetoresistance data in Fig.\ref{fig6}(c), the onset of electronic nematicity is around 15 K to 60 K depending on the magnetic field strength \cite{wenhaihu}. Hence, the electronic nematic transition seems to be separated from the CDW transition at least in  CsV$_3$Sb$_5$. More than that, the signature of this nematic transition can also be found in 
 $\mu$SR,   coherent phonon spectroscopy and Raman spectroscopy \cite{harter,wangnanlin,miaohu_xrd}. The muon spin relaxation rate has a second feature around  $T=$30K in addition to the onset of the primary TRSB CW transition \cite{yuli}. Optical data performing coherent phonon spectroscopy show a 3.1 THz peak appear below 30K $\sim$ 60K in addition to the 1.3 THz peak coupled to the onset of the CDW and 4.1 THz normal peaks \cite{harter,wangnanlin}, as shown in Fig.\ref{fig6}(d). 
Raman spectroscopy  also found additional peaks below 30 K \cite{miaohu_xrd}, as plotted in  Fig.\ref{fig6}(e). A similar 40K transition was also identified from the NMR measurement \cite{zhourui}.
Hence, it is highly possible that there is an electronic nematic transition around 30K $\sim$ 40K in CsV$_3$Sb$_5$. 

Additionally, STM experiments show an in-plane 1$\times$4  charge modulation below $50\sim60$ K \cite{hezhao,gaohongjun,zywang}, as shown in Fig.\ref{fig6}(f). 
From the Fourier transform of STM topographs shown in Fig.\ref{fig6}(g), there is one additional CDW peak  (Q$_{1Q}$)  appearing alongside the structural Bragg peaks (Q$_{Bragg}$) and 2$\times$2 CDW peaks (Q$_{3Q}$) \cite{hezhao}.
Since similar 1$\times$4  charge orders have been widely found in cuprates \cite{fradkin,tranquada,mesaros}, this 1$\times$4 charge order has attracted considerable attention.
To date, however, bulk measurements such as x-ray scattering and NMR have failed to confirm this 1$\times$4 order \cite{miaohu3}. As it depends on the cleaved surface environment \cite{wutao,wilson_quantum,miaohu_xrd,miaohu2}, this 1$\times$4  charge  order is most likely a surface manifestation of the intermediate 30-60K transition, which is supported by the DFT calculations \cite{yaoyugui}. We should note that observing diffuse quasi-1D correlations in a system that has three such domains is very challenging in conventional x-ray measurements, which calls for further exploration.

\section{Theories and Models}

\begin{figure*}
	\begin{center}
		\includegraphics[width=7.0in]{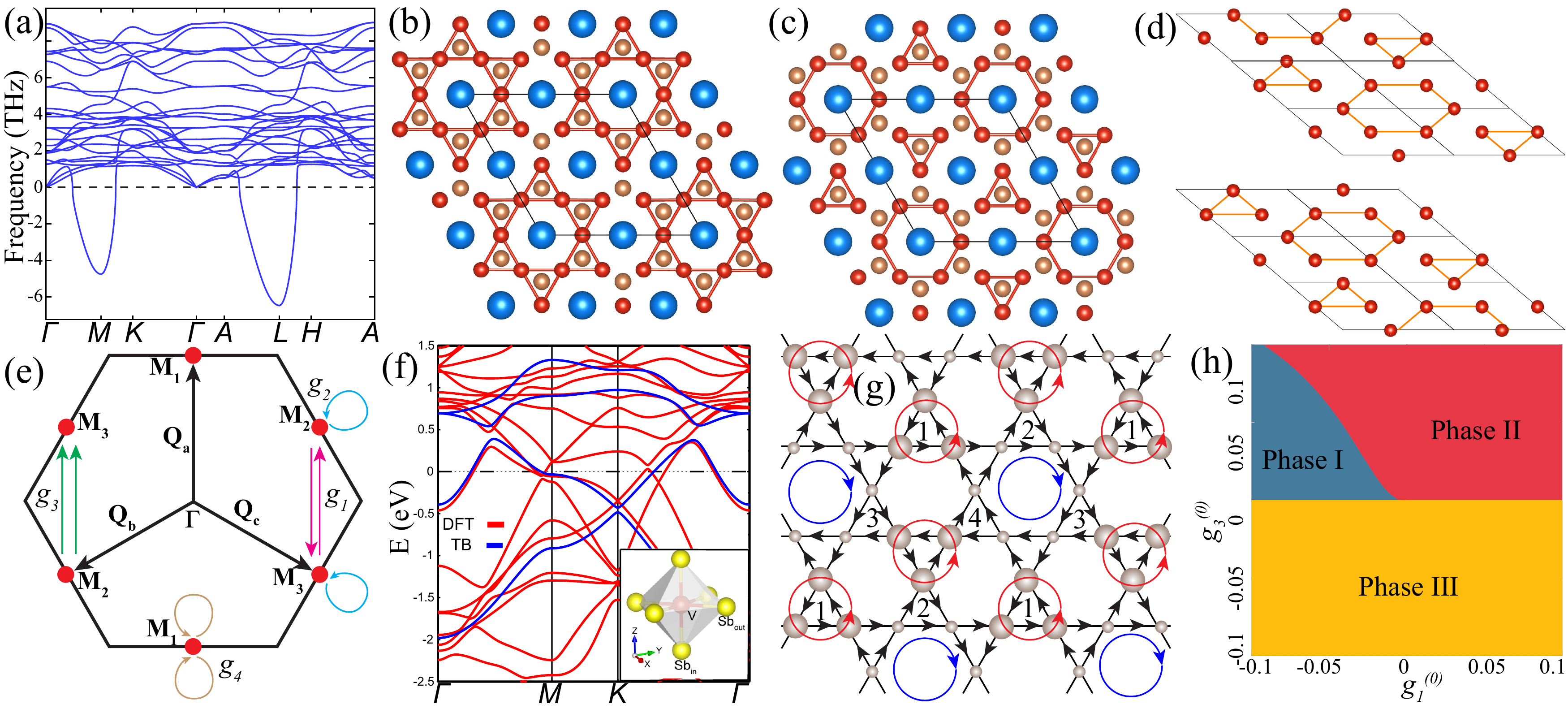}\caption{(a) Phonon spectrum calculated for CsV$_3$Sb$_5$.  (b) (c) Star of David and Tri-hexagonal distortions for CsV$_3$Sb$_5$ \cite{binghai}. (d) 3D structure distortion for AV$_3$Sb$_5$ with $\pi$ shift between the adjacent Kagome layers \cite{harter}. (e) The low-energy effective theory of three VH points $M_{1-3}$ for AV$_3$Sb$_5$ \cite{balents,fengxl2}. The arrows denote the scattering processes described by the interactions $g_{1-4}$.
		(f) Band structure for the minimal model for CsV$_3$Sb$_5$ \cite{guyuhao}. (g)  The flux configuration for the chiral flux phase \cite{feng}. (h) RG phase diagram for the effective model \cite{balents}.
			\label{fig7}}
	\end{center}
	\vskip-0.5cm
\end{figure*}

Theoretically, how one models and describes the AV$_3$Sb$_5$ materials, especially their unconventional CDW states, becomes a crucial question. 
As discussed above, DFT calculations qualitatively agree with the electronic structures of AV$_3$Sb$_5$ from ARPES measurements. Therefore, DFT calculations could provide a reasonable starting point for the understanding of AV$_3$Sb$_5$.
Since the structural transition is found to play a vital role in the CDW formation, the most stable structural distortion can be probed by DFT. For example, in CsV$_3$Sb$_5$, phonon dispersion relations are calculated from the \textit{ab initio} DFT calculations shown in Fig.\ref{fig7}(a) \cite{binghai}. From the phonon modes, one finds that there are two negative energy soft modes around the M and L points respectively. The structural instabilities led by these soft modes, the "Star of David" (SoD) and "Tri-Hexagonal" (TrH) structure configurations are proposed to be the likely candidates for CDW structures \cite{binghai,wilson_quantum,harter}, as illustrated in Fig.\ref{fig7}(b) and (c). Note that TrH is also named as "Inverse Star of David" in the literature. Based on XRD data, STM and quantum oscillation measurements, the TrH state is suggested to be the promising ground state configuration below $T_{CDW}$ in a single layer model. To accomplish the $2 \times 2 \times 2$ structure modulation, a $\pi$ shift between the adjacent Kagome layer TrH distortions is needed \cite{harter,miaohu2}, as illustrated in Fig.\ref{fig7}(d). On the other hand, recent studies have suggested that the average structure shows signatures of both TrH and SoD structures in the staggered layer sequence \cite{wilson_quantum}, which calls for further investigation.

Beyond the structural transition, a model that captures the electronic properties of AV$_3$Sb$_5$ is important.  DFT calculations and ARPES measurements show multiple bands cross the Fermi level \cite{ortiz20,ortiz19}. As discussed above, the in-plane Sb $p_z$ orbital forms one electron pocket around the $\Gamma$ point and the V $d$ orbitals form multiple FSs around the M points, as illustrated in Fig.\ref{fig7}(e) \cite{feng}. 
It is very difficult to capture such a complicated Fermi surface topography in a  simplified tight-binding model. However, the essential electronic structure of AV$_3$Sb$_5$ is widely believed to be dominated by the quasiparticles around the VH points based on the following facts. First, the VH points are  very close to the Fermi level as obtained from DFT calculations and ARPES measurements \cite{ortiz20,zhouxingjiang,zhangyan}.
Second, the quasiparticle interference
spectrum shows that the dominant scattering momenta are 3Q (Q$_{a}$, Q$_{b}$, Q$_{c}$) related to three M points as well as the $\Gamma$ point FS induced $q_1$ scattering \cite{zywang,hezhao}, as illustrated in Fig.\ref{fig7}(e).  Finally, the CDW gap size is maximum around the VH points while its vanishes at the $\Gamma$  pocket \cite{zhangyan,zhouxingjiang}. Therefore, a minimal model capturing the VH points and  $\Gamma$ point FS could faithfully describe the physics behind AV$_3$Sb$_5$ \cite{wilson_quantum}.
Following this spirit, a minimal 4 band model based on the V local $d_{X^2-Y^2}$ orbital and in-plane Sb $p_z$ orbital is proposed as shown in Fig.\ref{fig7}(f) \cite{guyuhao}. And the V local $d_{X^2-Y^2}$ orbital  model is adiabatically connected to the nearest-neighbor tight-binding model in the kagome lattice. This model provides a solid ground for further theoretical investigation.

The most intriguing property of the AV$_3$Sb$_5$ CDW is its TRSB. However, neutron scattering, NMR and $\mu$SR experiments have already ruled out the possibility of long-range magnetic order with conventional moments in the resolution of the measurements \cite{ortiz19,musr1,yuli,graf}. This feature is reminiscent of long-discussed flux phases in condensed matter, such as the Haldane model on the honeycomb lattice \cite{haldane}.  Moreover, the flux phases breaking TRSB are also widely discussed in cuprate superconductors after the seminal study by Affleck and Marston in t-J models \cite{affleck,lee}. Generalizing this idea, Varma proposed a loop-current phase formed in the Cu-O triangles \cite{varma} and Chakravarty et al. proposed the d-density wave state with staggered flux in Cu square plaquettes \cite{chakravarty}. Both states break the time-reversal symmetry and are candidates for the pseudogap in cuprates \cite{varma06,varma97,varma,chakravarty,norman_rev,keimer}. 

For kagome lattices and other hexagonal lattices, the 3Q electronic instabilities at VH filling have been widely discussed \cite{martin,taoli,kun,jxli,chubukov,qhwang13,thomale13,qhwang12,thomale12,motome}, including chiral spin density wave order, charge bond orders, intra-unit cell CDW and $d$+i$d$ SC etc. 
 Based on the minimal model and the 3Q electronic instabilities, several TRSB flux states have been proposed to explain the TRSB.  The most promising candidate is  the chiral flux phase among the 18 flux classes \cite{feng,titus,balents,linyp,fengxl2}. In this chiral flux state shown in Fig.\ref{fig7}(g), there are two special flux loops. The two anti-clockwise triangle current flux loops (red circles) form a honeycomb lattice and the clockwise hexagonal current flux (blue circle) forms a triangular lattice. The charge order of the chiral flux phase  coincides with $2 \times 2$ charge order and the TrH lattice configuration \cite{feng}.

Microscopically, how to stabilize the flux state is still under debate. Starting from the VH points, the low-energy effective theory of AV$_3$Sb$_5$ can be constructed by projection \cite{balents,fengxl2}, as illustrated in Fig.\ref{fig7}(g). Using the parquet renormalization group, various leading and subleading instabilities have been determined, including superconductivity, charge order, orbital moment, and spin density waves \cite{balents}. For example, a renormalization group phase diagram is shown in Fig.\ref{fig7}(h) when the bare interaction is $g_2>0$. There are three possible phases, Phase I, II and III. Although both the leading and subleading  instabilities have been discussed in this work, we only focus on the leading one. Among these three phases, the leading instability of  phase II is the ''imaginary charge density wave'' (iCDW), which is the low-energy version of the flux phase. 	In this case, we can find that the TRSB phase can be stabilized if the bare interaction $g_1$ is negative and $g_2$, $g_3$, $g_4$ are positive. But how to achieve attractive interactions needs to be further explored \cite{balents}. An extended Hubbard model with on-site Hubbard interaction U and the nearest-neighbor Coulomb interaction V is also proposed to stable the TRSB order \cite{titus,feng}. However, the TRSB order has not been found  in the realistic parameter region in this type of model. Phenomenologically, the various Ginzburg Landau theory approaches have also been discussed to describe the TRSB phases \cite{titus,linyp,balents}.

\section{Superconductivity}
Superconductivity  remains the important property of  the AV$_3$Sb$_5$ materials.  We will focus on discussing  superconducting mechanism and pairing symmetry.  Whether  SC is  driven by  electron-phonon coupling, or unconventionally driven by electron-electron correlation, is the central issue we need to address. To find clues for this hard-core question, we will first focus on the superconducting pairing symmetries of AV$_3$Sb$_5$.  Since the inversion symmetry ${\cal I}$ is always a good symmetry for  AV$_3$Sb$_5$ as found in SHG measurements \cite{yuli}, the spin singlet pairing and spin triplet pairing must be separated. 
 
\begin{figure*}
	\begin{center}
		\includegraphics[width=7.0in]{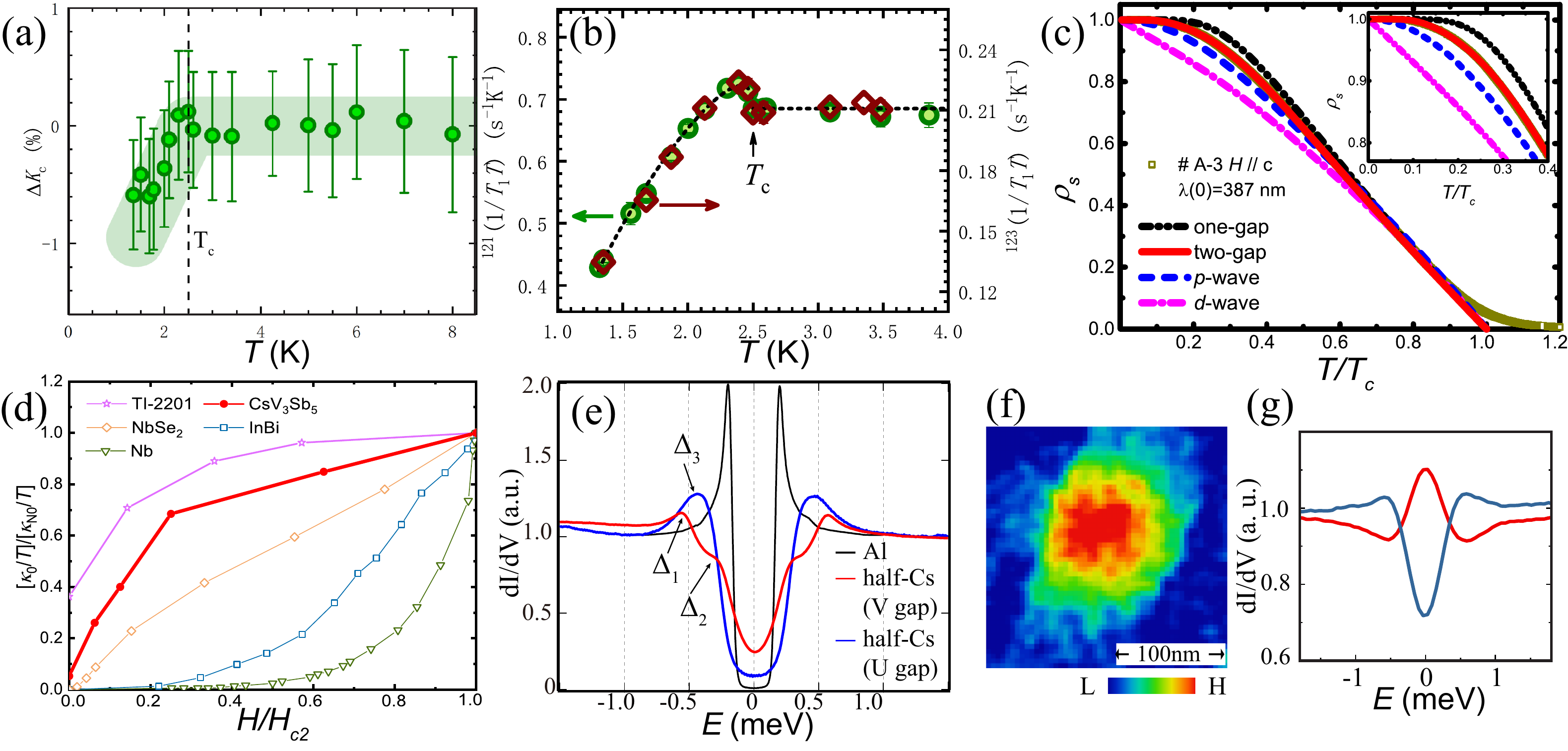}\caption{(a) Temperature dependence of  the Knight shift $\Delta K$ of $^{121}Sb$ for CsV$_3$Sb$_5$ with $H//c$ \cite{zhengli}.  (b) Temperature dependence of $^{121}(1/T_1T)$ (left axis) and $^{123}(1/T_1T)$  (right axis). A Hebel-Slichter coherence peak appears just below $T_c$ for CsV$_3$Sb$_5$ \cite{zhengli}. (c) The normalized superfluid density $\rho_s$ for CsV$_3$Sb$_5$ as a function of the reduced temperature $T/T_c$ \cite{yuanhq}. The dash-dot-dotted, solid, dashed, and dash-dotted lines respectively represent fits to models with a single s-wave gap, two s-wave gaps, a p-wave gap, and a d-wave gap. The inset zooms into the low temperature region. 			
			(d) The normalized residual linear term $\kappa_0/T$ of CsV$_3$Sb$_5$ as a function of $H/H_{c2}$. Similar data of $Nb$, InBi, NbSe$_2$ and an overdoped d-wave cuprate superconductor Tl-2201 are shown for comparison \cite{shiyanli}. (e) Two kinds of superconducting gap spectra observed on half Cs surface for CsV$_3$Sb$_5$ \cite{fengdonglai}. (f) $dI/dV$ map showing a superconducting vortex on the Cs surface for CsV$_3$Sb$_5$ \cite{zywang}. (g) Tunneling spectra obtained in the vortex core (red) with zero-bias peak and outside the vortex (dark-blue) \cite{zywang}.
			\label{fig8}}
	\end{center}
	\vskip-0.5cm
\end{figure*}

To reveal the pairing properties, multiple experimental techniques have been applied. The first task is to determine whether the Cooper pairs form a singlet or triplet, which can be determined through the temperature-dependent spin susceptibility. 
From the NMR spectrum shown in Fig.\ref{fig8}(a), one finds that the temperature dependent z-direction Knight shift of  $^{121}Sb$ drops below the SC transition $T_c$ in CsV$_3$Sb$_5$\cite{zhengli}. The Knight shift in the other two directions also show a similar dropping below $T_c$ \cite{zhengli}.
Therefore, the ground state of AV$_3$Sb$_5$ belongs to a spin-singlet SC. Additionally, the $\mu$SR measurements fail to detect any additional TRSB signals below $T_c$ comparing to the distinct increasing in the  Sr$_2$RuO$_4$ SC \cite{sr2ruo4}, suggesting a time-reversal invariant superconducting order parameter \cite{musr1,musr2,yuli}. Therefore, the SC order parameter of AV$_3$Sb$_5$ belongs to the time-reversal preserved spin singlet. 

The superconducting gap structure can also provide information about pairing symmetry. 
A Hebel-Slichter coherence peak appears just below $T_c$ in CsV$_3$Sb$_5$  from the spin-lattice relaxation measurement of the $^{121/123}$Sb nuclear quadrupole resonance (NQR) \cite{zhengli}, as shown in Fig.\ref{fig8}(b).  This coherence peak is widely known as a hallmark for a gapped conventional s-wave SC \cite{hebel1,hebel2}. Moreover, an exponential temperature dependence of magnetic penetration depth is found at low temperatures,  suggesting a nodeless superconducting gap structure for  CsV$_3$Sb$_5$ \cite{yuanhq,musr2}, as shown in Fig.\ref{fig8}(c). No sub-gap resonance state is found near non-magnetic impurities while the magnetic impurities destroy the SC quite efficiently from STM measurements \cite{fengdonglai}. Hence, the SC of  AV$_3$Sb$_5$  is most likely a conventional, fully gapped $s$-wave SC. This feature is also consistent with the weakly correlated nature of AV$_3$Sb$_5$ and remarkable electron-phonon coupling of the V-derived bands found from ARPES \cite{zhouxingjiang}.

However, this simple picture is complicated by experimental observation of nodes or deep minima in the superconducting gap. From thermal transport measurements, a finite residual thermal conductivity $\kappa_0$ at $T \rightarrow 0$ has been found in CsV$_3$Sb$_5$, which suggests a nodal feature of the pairing order parameter \cite{shiyanli,xxwu}.  This residual thermal conductivity $\kappa_0$  also shows a similar magnetic field dependence found in a $d$-wave cuprate, as shown in Fig.\ref{fig8}(d). Additionally, a multiple-gap feature is resolved from the mK-STM measurements,  as shown in Fig.\ref{fig8}(e).
 The multi-gap behavior agrees with the multiple FSs revealed from the DFT calculations and the ARPES measurement. Interestingly, in different regions of CsV$_3$Sb$_5$, both the U-shaped and V-shaped suppression of the density of states (DOS) have been observed at the Fermi level with a relatively large residual DOS that can hardly be explained by thermal excitations  \cite{fengdonglai,gaohongjun}. These findings, on the other hand, prefer a superconducting gap with nodes.

This leads to a seeming dichotomy between gapless excitations in the SC state and a conventionally gapped s-wave SC for AV$_3$Sb$_5$. However, if we take the TRSB normal states into account, the gapless excitations arise  within a fully-opened superconducting gap \cite{guyuhao}. There are two key discrete symmetries in SCs to guarantee the presence of Cooper pairing, time-reversal ${\cal T}$ and inversion symmetry ${\cal I}$ \cite{sigrist,anderson1,anderson2}.  For the even-parity spin-singlet pairing formed by $(c_{k,\uparrow}c_{-k,\downarrow}-c_{k,\downarrow}c_{-k,\uparrow})$,  the system at least contains time-reversal symmetry ${\cal T}$ because of the ${\cal T}$ maps a $|k,\uparrow>$ state to $|-k,\downarrow>$ state. Similarly, the odd-parity, spin-triplet pairing needs inversion symmetry ${\cal I}$ owing to the fact that $I$ maps a $|k,\uparrow>$ state to $|-k,\uparrow>$ state. These two symmetry conditions are known as Anderson's theorem \cite{sigrist,anderson1,anderson2}.  For AV$_3$Sb$_5$ SC cases,  the normal state before the SC transition breaks the ${\cal T}$ symmetry as discussed above. Therefore, the edge modes on CDW domain walls or other places where the TRSB dominates, cannot be gapped out by the SC pairing. These gapless excitations could contribute a finite residual thermal conductivity.

Although SC seems to be conventional, the nontrivial band structure of AV$_3$Sb$_5$ could lead to nontrivial excitations. Based on Fu-Kane's seminal proposal, if the helical Dirac surface state of a topological insulator are in proximity to an $s$-wave SC, Majorana zero modes (MZMs) may arise inside the vortex cores of the superconducting Dirac surface states \cite{fukane}. The proposal has been widely used in Bi$_2$Te$_3$/NbSe$_2$ heterostructures, and in the iron-based SC Fe(Te,Se), LiOHFeSe etc \cite{jfjia,yin_fe, haoning, zjwang,xxwu2,pzhang,gxu,hongding,lifese}. Similar to these aforementioned materials, AV$_3$Sb$_5$ hosts Dirac surface states near the Fermi energy \cite{ortiz20} that can open a superconducting gap below T$_c$. Therefore, MZMs are theorized to emerge inside the vortex core. Using STM, zero-bias states with spatial evolution similar to the zero-bias peaks in Bi$_2$Te$_3$/NbSe$_2$ heterostructures have been resolved in the vortex cores of CsV$_3$Sb$_5$ \cite{zywang}, as shown in Fig.\ref{fig8}(f,g).

\begin{figure*}
	\begin{center}
		\includegraphics[width=7.0in]{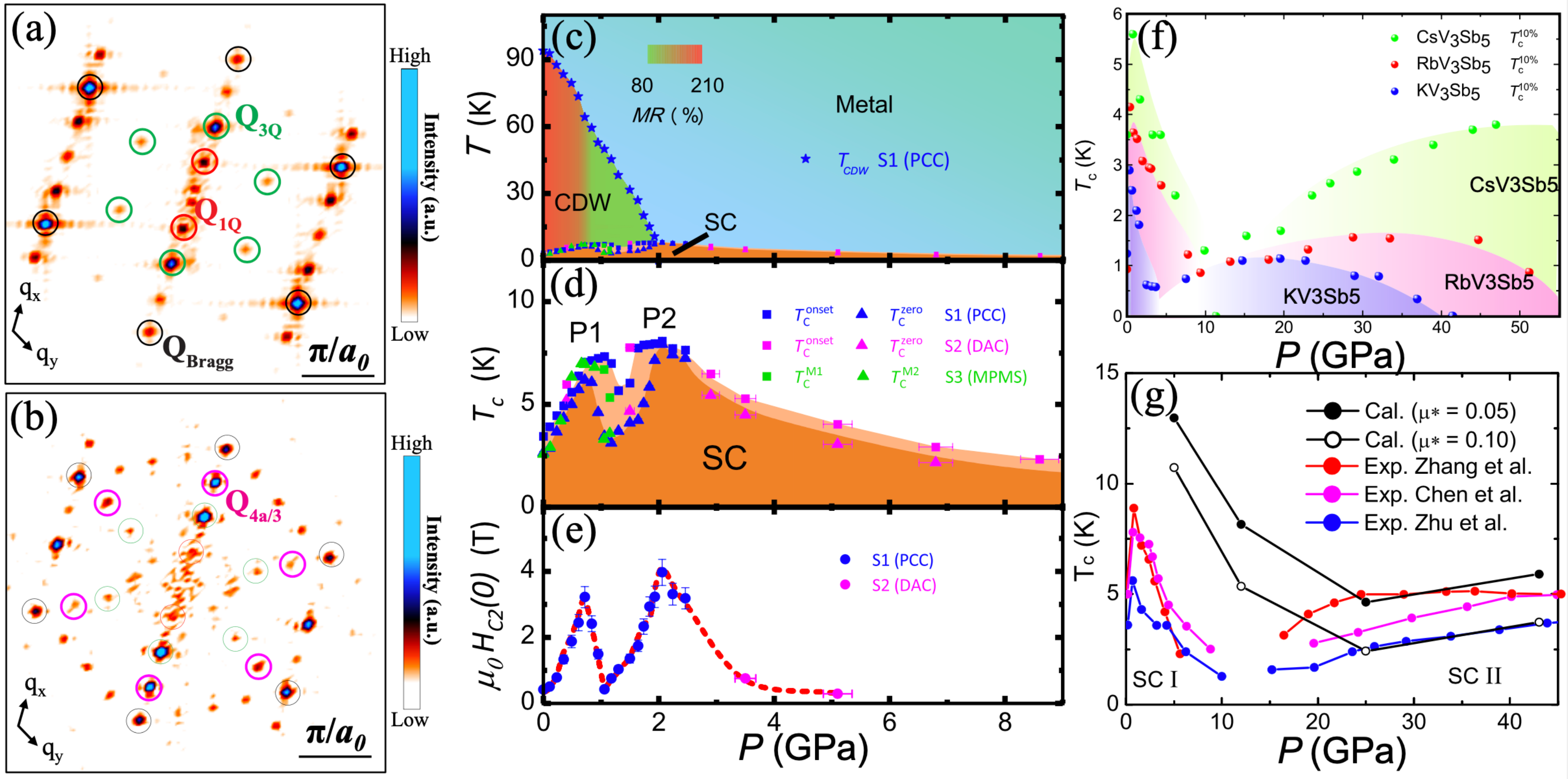}\caption{(a)  Fourier transformation of atomically-resolved STM topography of Sb surface for CsV$_3$Sb$_5$. (b) $dI/dV$ map at -0.25 meV for CsV$_3$Sb$_5$ at $T_{electron}$=300mK. Comparing to (a), there are additional peaks at $Q_{4/3a}$ \cite{gaohongjun}.  (c) Phase diagram for CsV$_3$Sb$_5$ with pressure. CDW transition temperature $T_{CDW}$ gradually suppressed with increasing the pressure. The color inside the CDW represents the magnitude of magnetoresistance measured at 9T and 10K. (d) Pressure dependence of superconducting transition temperatures showing two dome behavior. (e) Pressure dependence of upper critical field at zero temperature \cite{yingjianjun}.
			(f) Temperature–pressure phase diagram of AV$_3$Sb$_5$ \cite{lishiyan_pre}.
			(g) Electron-phonon calculated $T_c$ for CsV$_3$Sb$_5$ and its comparison with experiments \cite{luzy}.
			\label{fig9}}
	\end{center}
	\vskip-0.5cm
\end{figure*}

In addition, CsV$_3$Sb$_5$ may host an intriguing electronic state, known as the pair density wave (PDW), in which the Cooper-pair density modulates spatially at a characteristic wave vector. 
A low-temperature STM study on CsV$_3$Sb$_5$ found that both the height of superconducting coherence peak and the zero-energy gap-depth show spatial modulations with a distinct periodicity of $4a/3$, suggesting a PDW
 state \cite{gaohongjun}. In the Fourier transforms of the differential conductance maps taken inside the superconducting gap, six additional Q$_{4/3a}$ modulation peaks were found in addition to the 2$\times$2 CDW peaks Q$_{3Q}$, 1$\times$ 4 CDW peaks Q$_{1Q}$ and Bragg peaks shown in Fig.\ref{fig9}(a,b). Four of these additional Q$_{4/3a}$  vectors cannot be obtained by linear combinations of Q$_{3Q}$ and Q$_{1Q}$ peaks, which provides evidence for the PDW in AV$_3$Sb$_5$ \cite{gaohongjun}.

As the superconductivity in AV$_3$Sb$_5$ arises within the preexisting CDW states, exploring the correlation between these two states can help to reveal the underlying physics \cite{chengjg,yingjianjun,chengxiaolong,yangzhaorong,yuanhq_pre,lishiyan_pre}. By applying external pressure to CsV$_3$Sb$_5$, CDW order becomes destabilized quickly and vanishes at 2 GPa while the SC state shows a double-peak behavior with a maximum of 8K around 2 GPa \cite{yingjianjun,chengjg}, as plotted in Fig.\ref{fig9}(c). The competition between CDW and SC is a common feature for all AV$_3$Sb$_5$ materials, while the double-peak behavior is clearest in CsV$_3$Sb$_5$ \cite{lishiyan_pre}.
Hence, the CDW order highly correlates with SC in the low pressure region, known as SC I.  By further increasing pressure, a new SC dome, named SC II, appears for all AV$_3$Sb$_5$ material, as shown in Fig.\ref{fig9} (d). 
A recent DFT calculation with electron-phonon coupling shows that the $T_c$ calculated from the McMillan-Allen-Dynes formula qualitatively agrees with the experimental values obtained above the 20 GPa \cite{luzy}, as shown in Fig.\ref{fig9} (e). Hence, the SC II state at high pressure  likely stems from the electron-phonon coupling. However, the $T_c$ calculated based on electron-phonon coupling in the low pressure range is far above the experimental values, which cannot give rise to a reliable conclusion. The underlying pairing mechanism for AV$_3$Sb$_5$  needs more experimental exploration and theoretical analysis.

\section{Summary and Perspective}
In this article, we have reviewed the physical properties of  the newly discovered Kagome materials AV$_3$Sb$_5$. Owing to tremendous efforts during the past year, we have achieved considerable understandings of AV$_3$Sb$_5$, which can be summarized as  
\begin{itemize}
	\item AV$_3$Sb$_5$ is a quasi-2D electronic system with cylindrical Fermi surfaces, where the electronic properties are dominated by the V-Sb kagome layers. 
	\item AV$_3$Sb$_5$ is a multi-band system with at least four bands crossing the Fermi level. The FS around the $\Gamma$ point is attributed to the Sb $p_z$ bands while FSs around the BZ boundary mainly consisting of the V $d$ orbitals. The VH points at the M points play an important role in the unconventional properties of AV$_3$Sb$_5$. 
	\item Owing to band inversions at M points, the AV$_3$Sb$_5$ is a $Z_2$ topological metal with unconventional surface states.
	\item The correlation strength of AV$_3$Sb$_5$ is weak based on DFT calculations and ARPES measurements.
	\item AV$_3$Sb$_5$ undergoes a first-order phase transition into charge density wave order around 80 K to 104 K, depending on the A-site cation. Within the kagome layer, the CDW enlarges the unit cell to 2$\times$2 accompanied by a $c$-axis modulation.
	\item There is strong evidence for the emergence of time-reversal symmetry breaking inside the CDW state. Besides translational symmetry breaking and time-reversal symmetry breaking, inversion symmetry is persevered while the $C_6$ rotation symmetry is broken.
	\item The superconducting order parameter of  AV$_3$Sb$_5$ SC is a spin singlet with $T_c$ around 1-3K, depending on the A-site cation. SC appears to be a conventional s-wave with unconventional excitations inside the vortex core. CDW order is intertwined with SC in an unconventional way, inducing multiple SC domes under pressure. 
\end{itemize}

The discovery of AV$_3$Sb$_5$ SC opens a new route towards realizing unconventional orders within 2D kagome metals, which brings us a new platform to investigate the interplay between correlation, topology and geometric frustration. We hope that this review provides a broad picture of the recent progress on AV$_3$Sb$_5$ kagome materials and stimulates new research frontiers within kagome-related physics.

\section{Acknowledge}
We thank Hechang Lei, Hu Miao, Jianjun Ying, Xingjiang Zhou, Junfeng He, Shancai Wang, Li Yu, Xiaoli Dong, Fang Zhou, Yan Zhang, Nanling Wang, Huan Yang, Haihu Wen, He Zhao,  Ilija Zeljkovic, Binghai Yan, Ziqiang Wang, Zheng Li, Jianlin Luo, Yu Song, Huiqiu Yuan, Shiyan Li, Yajun Yan, Donglai Feng, Hui Chen, Geng Li, Hongjun Gao, Rui Zhou etc. for useful discussions. We also need thank Yuhao Gu, Yuxing Wang for the help of DFT calculations.
This work is supported by the Ministry of Science and Technology of China 973 program (Grant No. 2017YFA0303100), National Science Foundation of China (Grant No. NSFC-11888101) and the Strategic Priority Research Program of Chinese Academy of Sciences (Grant No. XDB28000000 and XDB33000000). K.J. acknowledges support from the start-up grant of IOP-CAS. SDW gratefully acknowledges support via the UC Santa Barbara NSF Quantum Foundry funded via the Q-AMASE-i program under award DMR-1906325. W.T. and X.H.Chen acknowledge support from  the National Key R$\&$D Program of the MOST of China (Grants No. 2017YFA0303000, 2016YFA0300201), the National Natural Science Foundation of China (Grants No. 11888101, 12034004), the Strategic Priority Research Program of Chinese Academy of Sciences (Grant No. XDB25000000), the Anhui Initiative in Quantum Information Technologies (Grant No. AHY160000), the Collaborative Innovation Program of Hefei Science Center, CAS (Grant No.2019HSC-CIP007). Z.Y.W. is supported by National Natural Science Foundation of China (No. 12074364). J. X. Yin and M. Hasan are supported by the Gordon and Betty Moore Foundation (GBMF4547 and GBMF9461).



\begin{thebibliography}{199}
	\bibitem{onsager}
	Onsager, Phys. Rev. {\bf65}, 117 (1944).
	
	\bibitem{huangkerson}
	Kerson Huang, {\it Statistical Mechanics}, John Wiley \& Sons, Inc. (1987).
	
	\bibitem{graphene1}
	A. H. Castro Neto, F. Guinea, N. M. R. Peres, K. S. Novoselov, A. K. Geim, 	Rev. Mod. Phys. {\bf81}, 109 (2009).
	
	\bibitem{graphene2}
	P. R. Wallace , Phys. Rev. {\bf71}, 622 (1947).
	
	\bibitem{geim}
	K. S. Novoselov et al., Science {\bf306}, 666 (2004).
	
	\bibitem{syozi}
	I. Syozi, Progress of Theoretical Physics {\bf6}, 306 (1951).
	
	\bibitem{mekata}
	M. Mekata, Physics Today {\bf56}, 12 (2003).
	
	\bibitem{yizhou}
	Yi Zhou, Kazushi Kanoda, and Tai-Kai Ng, Rev. Mod. Phys. {\bf89}, 025003  (2017).
	
	\bibitem{balents_rev}
	L. Balents, Nature {\bf464}, 199 (2010).
	
	\bibitem{norman1}
	M. R. Norman, Rev. Mod. Phys. {\bf 88}, 041002 (2016).
	
	\bibitem{norman2}
	C. Broholm, R. J. Cava, S. A. Kivelson, D. G. Nocera, M. R. Norman, T. Senthil, Science {\bf 367}, 263 (2020).
	
	\bibitem{villain}
	J. Villain,  R. Bidaux, J.-P. Carton, and R. Conte, J. Phys. France {\bf41}, 1263 (1980).
	
	
	\bibitem{qsl1}
	Ying Ran, Michael Hermele, Patrick A. Lee, and Xiao-Gang Wen, Phys. Rev. Lett. {\bf98}, 117205 (2007).
	
	\bibitem{qsl2}
	J. S. Helton, K. Matan, M. P. Shores, E. A. Nytko, B. M. Bartlett, Y. Yoshida, Y. Takano, A. Suslov, Y. Qiu, J.-H. Chung, D. G. Nocera, and Y. S. Lee, Phys. Rev. Lett. {\bf98}, 107204 (2007).
	
	\bibitem{jianghc}
	H. C. Jiang, Z. Y. Weng, and D. N. Sheng, Phys. Rev. Lett. {\bf101}, 117203 (2008).
	
	\bibitem{white}
	Simeng Yan, David A. Huse, Steven R. White,	Science {\bf332} 1173 (2011).
	
	\bibitem{schollwock}
	Stefan Depenbrock, Ian P. McCulloch, and Ulrich Schollwock, Phys. Rev. Lett. {\bf109}, 067201 (2012).
	
	\bibitem{heyc}
	Yin-Chen He, Michael P. Zaletel, Masaki Oshikawa, and Frank Pollmann, Phys. Rev. X {\bf7}, 031020 (2017).
	
	\bibitem{xiangtao}
	H. J. Liao et al., Phys. Rev. Lett. {\bf118}, 137202 (2017).
	
	\bibitem{sugang}
	Xi Chen, Shi-Ju Ran, Tao Liu, Cheng Peng, Yi-Zhen Huang, Gang Su, Sci. Bull. {\bf63}, 1545 (2018).
	
	
	\bibitem{yin_review}
	Jia-Xin Yin, Shuheng H. Pan and M. Zahid Hasan, Nat. Rev. Phys. {\bf3}, 249 (2021). 
	
	
	\bibitem{yin20}
	J.-X. Yin, Wenlong Ma, Tyler A. Cochran et al., Quantum-limit Chern topological magnetism in TbMn$_6$Sn$_6$, Nature {\bf583} 533 (2020).
	
	
	\bibitem{ye}
	L. Ye, Mingu Kang, Junwei Liu, et al., Massive Dirac fermions in a ferromagnetic kagome metal, Nature {\bf555}  638 (2018).
	
	\bibitem{yin18}
	J.-X. Yin, Songtian S. Zhang, Hang Li, et al., Giant and anisotropic many-body spin–orbit tunability in a strongly correlated kagome magnet, Nature {\bf562}, 91 (2018).
	
	\bibitem{nagaosa}
	Kenya Ohgushi, Shuichi Murakami, and Naoto Nagaosa, Spin anisotropy and quantum Hall effect in the kagome lattice: Chiral spin state based on a ferromagnet, Phys. Rev. B {\bf62}, R6065(R) (2000).
	
	\bibitem{franz}
	H.-M. Guo, M. Franz, Topological insulator on the kagome lattice, Phys. Rev. B {\bf80} 113102 (2009).
	
	
	\bibitem{fqh}
	Evelyn Tang, Jia-Wei Mei, and Xiao-Gang Wen, High-temperature fractional quantum Hall states, Phys. Rev. Lett. {\bf106}, 236802 (2011).
	
	\bibitem{flatband1}
	Zhiyong Lin, Jin-Ho Choi, Qiang Zhang, et al., Flatbands and emergent ferromagnetic ordering in Fe$_3$Sn$_2$	Kagome lattices, Phys. Rev. Lett. {\bf121}, 096401 (2018).
	
	\bibitem{flatband2}
	J. -X. Yin, Songtian S. Zhang, Guoqing Chang, et al. Negative flat band magnetism in a spin–orbit-coupled correlated kagome magnet, Nat. Phys. {\bf15}, 443 (2019).
	
	\bibitem{flatband3}
	Mingu Kang, Linda Ye, Shiang Fang et al., Dirac fermions and flat bands in the ideal kagome metal FeSn, Nat. Mat. {\bf19},163 (2020).
	
	\bibitem{yangbj}
	Jun-Won Rhim, Kyoo Kim and Bohm-Jung Yang, Nature {\bf584}, 59 (2020).
	
	\bibitem{cosn1}
	Zhonghao Liu et al., Nat. Comm. {\bf11}, 4002 (2020).
	
	\bibitem{cosn2}
	J.-X. Yin et al., Nat. Comm. {\bf11}, 4003 (2020).
	
	
	\bibitem{jxli}
	Shun-Li Yu and Jian-Xin Li, Phys. Rev. B {\bf85}, 144402 (2012).
	
	\bibitem{wenxg}
	Wing-Ho Ko, Patrick A. Lee, Xiao-Gang Wen, Phys. Rev. B {\bf79}, 214502 (2009)
	
	\bibitem{qhwang13}
	Wan-Sheng Wang, Zheng-Zhao Li, Yuan-Yuan Xiang, and Qiang-Hua Wang, Phys. Rev. B {\bf87}, 115135 (2013).
	
	\bibitem{thomale13}
	Maximilian L. Kiesel, Christian Platt, and Ronny Thomale, Phys. Rev. Lett. {\bf110}, 126405 (2013).
	
	
	
	
	\bibitem{ortiz19}
	B. R. Ortiz et al., Phys. Rev. Mater. {\bf3}, 094407 (2019).
	
	\bibitem{ortiz20}
	B. R.  Ortiz et al., Phys. Rev. Lett. {\bf125}, 247002 (2020). 
	
	\bibitem{note}
	Throughout this paper, the SC transition temperature $T_c$ is defined as the temperature where the resistivity goes to zero.
	
	
	\bibitem{ortiz21}
	Brenden R. Ortiz, Paul M. Sarte, Eric M. Kenney, Michael J. Graf, Samuel M. L. Teicher, Ram Seshadri, and Stephen D. Wilson, Phys. Rev. Materials {\bf5}, 034801 (2021).
	
	\bibitem{lei}
	Qiangwei Yin, Zhijun Tu, Chunsheng Gong, Yang Fu, Shaohua Yan, Hechang Lei, Chin. Phys. Lett. {\bf38}, 037403 (2021).
	
	\bibitem{shiyanli}
	C. C. Zhao  et al., arXiv:2102.08356
	
	
	
	
	\bibitem{hall20}
	S.-Y.  Yang et al., Sci. Adv. {\bf6}, eabb6003 (2020).
	
	\bibitem{edgecurrent}
	Y. Wang  et al., arXiv:2012.05898.
	
	\bibitem{graf}
	Eric M. Kenney, Brenden R. Ortiz, Chennan Wang, Stephen D. Wilson, Michael J. Graf, J. Phys.: Condens. Matter {\bf33} 235801.
	
	\bibitem{topocdw}
	Y.-X. Jiang  et al., arXiv:2012.15709. Nat. Mater. (2021). 
	
	\bibitem{xhchen}
	F. H. Yu, T. Wu, Z. Y. Wang, B. Lei, W. Z. Zhuo, J. J. Ying, X. H. Chen, Phys. Rev. B {\bf104}, 041103 (2021).
	
	\bibitem{wenhaihu}	
	Ying Xiang, Qing Li, Yongkai Li, Wei Xie, Huan Yang, Zhiwei Wang, Yugui Yao, Hai-Hu Wen, arXiv:2104.06909.
	
	\bibitem{dongxiaoli}	
	Shunli Ni, Sheng Ma, Yuhang Zhang, Jie Yuan, Haitao Yang, Zouyouwei Lu, Ningning Wang, Jianping Sun, Zhen Zhao, Dong Li, Shaobo Liu, Hua Zhang, Hui Chen, Kui Jin, Jinguang Cheng, Li Yu, Fang Zhou, Xiaoli Dong, Jiangping Hu, Hong-Jun Gao, Zhongxian Zhao, Chinese Phys. Lett. {\bf38} 057403 (2021).	
	
	
	
	\bibitem{miaohu_xrd}
	H. X. Li, T. T. Zhang, Y. Y. Pai, C. Marvinney, A. Said, T. Yilmaz, Q. Yin, C. Gong, Z. Tu, E. Vescovo, R. G. Moore, S. Murakami, H. C. Lei, H. N. Lee, B. Lawrie, and H. Miao, Phys. Rev. X {\bf11}, 031050 (2021).
	
	
	\bibitem{miaohu2}
	H. Miao, H. X. Li, H. N. Lee, A. Said, H. C. Lei, J. X. Yin, M. Z. Hasan, Ziqiang Wang, Hengxin Tan, Binghai Yan, 	arXiv:2106.10150.
	
	
	\bibitem{sato}
	Kosuke Nakayama, Yongkai Li, Min Liu, Zhiwei Wang, Takashi Takahashi, Yugui Yao, and Takafumi Sato, arXiv:2104.08042.
	
	\bibitem{zhangyan}
	Zhengguo Wang, Sheng Ma, Yuhang Zhang, Haitao Yang, Zhen Zhao, Yi Ou, Yu Zhu, Shunli Ni, Zouyouwei Lu, Hui Chen, Kun Jiang, Li Yu, Yan Zhang, Xiaoli Dong, Jiangping Hu, Hong- Jun Gao, and Zhongxian Zhao, arXiv:2104.05556.
	
	\bibitem{wangshancai}
	Zhonghao Liu, Ningning Zhao, Qiangwei Yin, Chunsheng Gong, Zhijun Tu, Man Li, Wenhua Song, Zhengtai Liu, Dawei Shen, Yaobo Huang, Kai Liu, Hechang Lei, and Shancai Wang, arXiv:2104.01125.
	
	\bibitem{shiming}
	Yong Hu, Samuel M. L. Teicher, Brenden R. Ortiz, Yang Luo, Shuting Peng, Linwei Huai, J. Z. Ma, N. C. Plumb, Stephen D. Wilson, J. F. He, and M. Shi, arXiv:2104.12725.
	
	\bibitem{comin}
	Mingu Kang, Shiang Fang, Jeong-Kyu Kim, Brenden R. Ortiz, Jonggyu Yoo, Byeong-Gyu Park, Stephen D. Wilson, Jae-Hoon Park, and Riccardo Comin, arXiv:2105.01689.
	
	\bibitem{shendawei}
	Soohyun Cho, Haiyang Ma, Wei Xia, Yichen Yang, Zhengtai Liu, Zhe Huang, Zhicheng Jiang, Xiangle Lu, Jishan Liu, Zhonghao Liu, Jinfeng Jia, Yanfeng Guo, Jianpeng Liu, and Dawei Shen, arXiv:2105.05117.
	
	\bibitem{hejunfeng}
	Yang Luo, Shuting Peng, Samuel M. L. Teicher, Linwei Huai, Yong Hu, Brenden R. Ortiz, Zhiyuan Wei, Jianchang Shen, Zhipeng Ou, Bingqian Wang, Yu Miao, Mingyao Guo, M. Shi, Stephen D. Wilson, and J. F. He, arXiv:2106.01248.
	
	\bibitem{borisenko}
	Rui Lou, Alexander Fedorov, Qiangwei Yin, Andrii Kuibarov, Zhijun Tu, Chunsheng Gong, Eike F. Schwier, Bernd Bchner, Hechang Lei, and Sergey Borisenko, arXiv:2106.06497.
	
	\bibitem{zhouxingjiang}
	Hailan Luo, Qiang Gao, Hongxiong Liu, Yuhao Gu, Dingsong Wu, Changjiang Yi, Junjie Jia, Shilong Wu, Xiangyu Luo, Yu Xu, Lin Zhao, Qingyan Wang, Hanqing Mao, Guodong Liu, Zhihai Zhu, Youguo Shi, Kun Jiang, Jiangping Hu, Zuyan Xu, X. J. Zhou, arXiv:2107.02688.
	
	\bibitem{wilson_quantum}
	Brenden R. Ortiz, Samuel M. L. Teicher, Linus Kautzsch, Paul M. Sarte, Jacob P. C. Ruff, Ram Seshadri, Stephen D. Wilson, arXiv:2104.07230.
	
	
	\bibitem{hussey}
	N. E. Hussey, M. Abdel-Jawad, A. Carrington, A. P. Mackenzie and L. Balicas, Nature {\bf425}, 814 (2003). 
	
	\bibitem{keimer}
	B. Keimer, S. A. Kivelson, M. R. Norman, S. Uchida and J. Zaanen, Nature {\bf518}, 179 (2015).
	
	\bibitem{iron_book}
	N. L. Wang, H. Hosono, and P. Dai, {\it Iron-Based Superconductors: Materials, Properties and Mechanisms} (Taylor and Francis, London, New York, 2012).
	
	\bibitem{wangyl}	
	Jianzhou Zhao, Weikang Wu, Yilin Wang, Shengyuan A. Yang, Phys. Rev. B {\bf103}, 241117 (2021).	
	
	\bibitem{z2}
	L. Fu and C. L. Kane, Phys. Rev. B {\bf76}, 045302 (2007).
	
	
	
	
	\bibitem{hezhao}
	He Zhao, Hong Li, Brenden R. Ortiz, Samuel M. L. Teicher, Taka Park, Mengxing Ye, Ziqiang Wang, Leon Balents, Stephen D. Wilson, Ilija Zeljkovic, arXiv:2103.03118.
	
	\bibitem{zywang}
	Zuowei Liang, Xingyuan Hou, Wanru Ma, Fan Zhang, Ping Wu, Zongyuan Zhang, Fanghang Yu, J. -J. Ying, Kun Jiang, Lei Shan, Zhenyu Wang, X. -H. Chen, Phys. Rev. X {\bf11}, 031026 (2021).
	
	\bibitem{gaohongjun}
	Hui Chen, Haitao Yang, Bin Hu, Zhen Zhao, Jie Yuan, Yuqing Xing, Guojian Qian, Zihao Huang, Geng Li, Yuhan Ye, Qiangwei Yin, Chunsheng Gong, Zhijun Tu, Hechang Lei, Shen Ma, Hua Zhang, Shunli Ni, Hengxin Tan, Chengmin Shen, Xiaoli Dong, Binghai Yan, Ziqiang Wang, Hong-Jun Gao, arXiv:2103.09188.	
	
	\bibitem{fengdonglai}	
	Han-Shu Xu, Ya-Jun Yan, Ruotong Yin, Wei Xia, Shijie Fang, Ziyuan Chen, Yuanji Li, Wenqi Yang, Yanfeng Guo, Dong-Lai Feng, arXiv:2104.08810.		
	
	\bibitem{lihong}	
	Hong Li, He Zhao, Brenden R. Ortiz, Takamori Park, Mengxing Ye, Leon Balents, Ziqiang Wang, Stephen D. Wilson, Ilija Zeljkovic, arXiv:2104.08209.	
	
	\bibitem{shumiya}	
	Nana Shumiya et al., Intrinsic nature of chiral charge order in the kagome superconductor RbV3Sb5, Phys. Rev. B {\bf104}, 035131 (2021).	
	
	\bibitem{yaoyugui}	
	Zhiwei Wang, Yu-Xiao Jiang, Jia-Xin Yin, Yongkai Li, Guan-Yong Wang, Hai-Li Huang, Shen Shao, Jinjin Liu, Peng Zhu, Nana Shumiya, Md Shafayat Hossain, Hongxiong Liu, Youguo Shi, Junxi Duan, Xiang Li, Guoqing Chang, Pengcheng Dai, Zijin Ye, Gang Xu, Yanchao Wang, Hao Zheng, Jinfeng Jia, M. Zahid Hasan, Yugui Yao, 	Phys. Rev. B {\bf104}, 075148 (2021).
	
	\bibitem{wutao}	
	D. W. Song, L. X. Zheng, F. H. Yu, J. Li, L. P. Nie, M. Shan, D. Zhao, S. J. Li, B. L. Kang, Z. M. Wu, Y. B. Zhou, K. L. Sun, K. Liu, X. G. Luo, Z. Y. Wang, J. J. Ying, X. G. Wan, T. Wu, X. H. Chen, 	arXiv:2104.09173.		
	

	
	
	
	\bibitem{musr1}
	C. Mielke III, D. Das, J.-X. Yin, H. Liu, R. Gupta, C.N. Wang, Y.-X. Jiang, M. Medarde, X. Wu, H.C. Lei, J.J. Chang, P. Dai, Q. Si, H. Miao, R. Thomale, T. Neupert, Y. Shi, R. Khasanov, M.Z. Hasan, H. Luetkens, Z. Guguchia, arXiv:2106.13443.
	
	\bibitem{yuli}	
	Li Yu, Chennan Wang, Yuhang Zhang, Mathias Sander, Shunli Ni, Zouyouwei Lu, Sheng Ma, Zhengguo Wang, Zhen Zhao, Hui Chen, Kun Jiang, Yan Zhang, Haitao Yang, Fang Zhou, Xiaoli Dong, Steven L. Johnson, Michael J. Graf, Jiangping Hu, Hong-Jun Gao, Zhongxian Zhao, arXiv:2107.10714.	
	
	\bibitem{nagaosa_ahe}	
	Naoto Nagaosa, Jairo Sinova, Shigeki Onoda, A. H. MacDonald, and N. P. Ong, Rev. Mod. Phys. {\bf82}, 1539 (2010).	
	
	\bibitem{fengxl2}		
	Xilin Feng, Yi Zhang, Kun Jiang, Jiangping Hu, arXiv:2106.04395.		
	
	\bibitem{nematic1}	
	S. A. Kivelson, E. Fradkin, V. J. Emery, Nature {\bf 393}, 550 (1998).
	
	\bibitem{nematic2}
	Eduardo Fradkin, Steven A. Kivelson, Michael J. Lawler, James P. Eisenstein, Andrew P. Mackenzie, Annu. Rev. Condens. Matter Phys. {\bf1}, 153 (2010).
	
	\bibitem{wangnanlin}	
	Z. X. Wang, Q. Wu, Q. W. Yin, Z. J. Tu, C. S. Gong, T. Lin, Q. M. Liu, L. Y. Shi, S. J. Zhang, D. Wu, H. C. Lei, T. Dong, N. L. Wang, 	arXiv:2105.11393.
	
	\bibitem{harter}	
	Noah Ratcliff, Lily Hallett, Brenden R. Ortiz, Stephen D. Wilson, John W. Harter, arXiv:2104.10138.	
	
	\bibitem{zhourui}	
	J. Luo, Z. Zhao, Y. Z. Zhou, J. Yang, A. F. Fang, H. T. Yang, H. J. Gao, R. Zhou, Guo-qing Zheng, 	arXiv:2108.10263.
	
	
	\bibitem{fradkin}	
	Eduardo Fradkin, Steven A. Kivelson, and John M. Tranquada, Rev. Mod. Phys. {\bf87}, 457  (2015).
	
	\bibitem{tranquada}	
	M. Hucker, M. v. Zimmermann, G. D. Gu, Z. J. Xu, J. S. Wen, Guangyong Xu, H. J. Kang, A. Zheludev, and J. M. Tranquada, Phys. Rev. B {\bf83}, 104506 (2011).	
	
	\bibitem{mesaros}	
	A. Mesaros, K. Fujita, S. D. Edkins, M. H. Hamidian, H. Eisaki, S. Uchida, J. C. Séamus Davis, M. J. Lawler, Eun-Ah Kim, Proc. Natl. Acad. Sci. U.S.A., {\bf113}, 12661 (2016).
	
	\bibitem{miaohu3}	
	Haoxiang Li, Yu-Xiao Jiang, J. X. Yin, Sangmoon Yoon, Andrew R. Lupini, C. Nelson, A. Said, Y. M. Yang, H. C. Lei, Binghai Yan, Ziqiang Wang, M. Z. Hasan, H. N. Lee, H. Miao, arXiv:2109.03418.
	
	
	
	
	
	\bibitem{binghai}
	Hengxin Tan, Yizhou Liu, Ziqiang Wang, Binghai Yan, Phys. Rev. Lett. {\bf127}, 046401 (2021).
	
	\bibitem{feng}
	Xilin Feng, Kun Jiang, Ziqiang Wang, Jiangping Hu, Sci. Bull. {\bf66}, 1384 (2021).
	
	\bibitem{guyuhao}
	Yuhao Gu, Yi Zhang, Xilin Feng, Kun Jiang, Jiangping Hu, arXiv:2108.04703.
	
	
	\bibitem{haldane}
	F. D. M. Haldane, Phys. Rev. Lett. 61, 2015 (1988). 
	
	\bibitem{affleck}
	Ian Affleck and J. Brad Marston, Phys. Rev. B 37, 3774(R) (1998).
	
	\bibitem{lee}
	Menke U. Ubbens and Patrick A. Lee, Phys. Rev. B 46, 8434 (1992).
	
	\bibitem{varma}
	C. M. Varma, Phys. Rev. B {\bf55}, 14554(1997).
	
	\bibitem{chakravarty}
	Sudip Chakravarty, R. B. Laughlin, Dirk K. Morr, and Chetan Nayak, Phys. Rev. B {\bf63}, 094503 (2001).
	
	\bibitem{varma97}
	C. M. Varma, Phys. Rev. B {\bf55}, 14554(1997).
	\bibitem{varma06}
	C. M. Varma, Phys.Rev.B {\bf73}, 155113 (2006).
	
	\bibitem{norman_rev}
	M. R. Norman ,D. Pines, and C. Kallin,  Adv. Phys. {\bf54}, 715 (2005).
	
	
	\bibitem{martin}
	Ivar Martin and C. D. Batista, Phys. Rev. Lett. {\bf101}, 156402 (2008).
	
	\bibitem{taoli}
	Tao Li, EPL {\bf97} 37001 (2012).
	
	\bibitem{motome}
	Satoru Hayami and Yukitoshi Motome, Phys. Rev. B {\bf90}, 060402(R) (2014).
	
	\bibitem{kun}
	Kun Jiang, Yi Zhang, Sen Zhou, and Ziqiang Wang, Phys. Rev. Lett. {\bf114}, 216402 (2015).
	
	\bibitem{chubukov}
	R. Nandkishore, L. Levitov, and A. Chubukov, Nat. Phys. {\bf8}, 158 (2012).
	
	\bibitem{qhwang12}
	Wan-Sheng Wang, Yuan-Yuan Xiang, Qiang-Hua Wang, Fa Wang, Fan Yang, and Dung-Hai Lee, Phys. Rev. B {\bf85}, 035414 (2012).
	
	\bibitem{thomale12}
	Maximilian L. Kiesel, Christian Platt, Werner Hanke, Dmitry A. Abanin, and Ronny Thomale, Phys. Rev. B {\bf86}, 020507(R) (2012).
	
	
	\bibitem{titus} 
	M. Michael Denner, Ronny Thomale and Titus Neupert, arXiv:2103.14045.
	
	\bibitem{linyp}
	Yu-Ping Lin and Rahul M. Nandkishore, Phys. Rev. B {\bf104}, 045122 (2021).
	
	\bibitem{balents}
	Takamori Park, Mengxing Ye, Leon Balents, 	Phys. Rev. B {\bf104}, 035142 (2021).
	
	
	\bibitem{zhengli}	
	Chao Mu, Qiangwei Yin, Zhijun Tu, Chunsheng Gong, Hechang Lei, Zheng Li, Jianlin Luo, Chin. Phys. Lett.  {\bf38}, 077402 (2021).
	
	\bibitem{sr2ruo4}
	G. M. Luke, Y. Fudamoto, K. M. Kojima, M. I. Larkin, J. Merrin, B. Nachumi, Y. J. Uemura, Y. Maeno, Z. Q. Mao, Y. Mori, H. Nakamura and M. Sigrist, Nature {\bf394}, 558 (1998). 
	
	\bibitem{musr2}
	Ritu Gupta, Debarchan Das, Charles Hillis Mielke III, Zurab Guguchia, Toni Shiroka, Christopher Baines, Marek Bartkowiak, Hubertus Luetkens, Rustem Khasanov, Qiangwei Yin, Zhijun Tu, Chunsheng Gong, Hechang Lei, 	arXiv:2108.01574.
	
	
	
	
	\bibitem{hebel1}	
	L. C. Hebel and C. P. Slichter, Nuclear relaxation in superconducting aluminum, Phys. Rev. {\bf107}, 901 (1957).
	
	\bibitem{hebel2}		
	L. C. Hebel and C. P. Slichter, Nuclear spin relaxation in normal and superconducting aluminum, Phys. Rev. {\bf113}, 1504 (1959).
	
	\bibitem{yuanhq}	
	Weiyin Duan, Zhiyong Nie, Shuaishuai Luo, Fanghang Yu, Brenden R. Ortiz, Lichang Yin, Hang Su, Feng Du, An Wang, Ye Chen, Xin Lu, Jianjun Ying, Stephen D. Wilson, Xianhui Chen, Yu Song, Huiqiu Yuan, 
	Sci. China-Phys. Mech. Astron. {\bf64}, 107462 (2021).
	
	\bibitem{xxwu}
	Xianxin Wu et al., arXiv:2104.05671.
	
	\bibitem{anderson1}	
	P.W. Anderson, J. Phys. Chem. Solids {\bf11}, 26 (1959).
	
	\bibitem{anderson2}	
	P.W. Anderson, Phys. Rev. B {\bf30}, 4000 (1984).
	
	\bibitem{sigrist}	
	M. Sigrist, AIP Conf. Proc. {\bf1162}, 55 (2009).
	
	
	
	\bibitem{fukane}
	Liang Fu, and C. L. Kane, Phys. Rev. Lett. {\bf100}, 096407 (2008).
	
	\bibitem{jfjia}
	Jin-Peng Xu et al., Phys. Rev. Lett. {\bf114} 017001 (2015).
	
	\bibitem{yin_fe}
	J. X. Yin et al., Nat. Phys. {\bf11}, 543 (2015).
	
	\bibitem{zjwang}
	Z. J. Wang et al., Phys. Rev. B {\bf92}, 115119 (2015).
	
	\bibitem{xxwu2}
	X. X. Wu et al., Phys. Rev. B {\bf93}, 115129 (2016).
	
	\bibitem{pzhang}
	P. Zhang et al. Science {\bf360}, 182 (2018).
	
	\bibitem{hongding}
	D. F. Wang et al., Science {\bf362}, 333 (2018).
	
	\bibitem{gxu}
	Gang Xu, et al., Phys. Rev. Lett. {\bf117}, 047001 (2016).
	
	\bibitem{lifese}
	Qin Liu, Chen Chen, Tong Zhang, Rui Peng, Ya-Jun Yan, Chen-Hao-Ping Wen, Xia Lou, Yu-Long Huang, Jin-Peng Tian, Xiao-Li Dong, Guang-Wei Wang, Wei-Cheng Bao, Qiang-Hua Wang, Zhi-Ping Yin, Zhong-Xian Zhao, and Dong-Lai Feng, Phys. Rev. X {\bf8}, 041056 (2018).
	
	\bibitem{haoning}
	Ning Hao, Jiangping Hu, Nat. Sci. Rev. {\bf 6}, 213 (2019).
	
	
	\bibitem{yingjianjun}
	F. H. Yu, D. H. Ma, W. Z. Zhuo, S. Q. Liu, X. K. Wen, B. Lei, J. J. Ying, X. H. Chen, Nat. Comm. {\bf12}, 3645 (2021).
	
	\bibitem{chengjg}
	K. Y. Chen, N. N. Wang, Q. W. Yin, Z. J. Tu, C. S. Gong, J. P. Sun, H. C. Lei, Y. Uwatoko, J.-G. Cheng, Phys. Rev. Lett. {\bf126}, 247001 (2021).
	
	
	\bibitem{yuanhq_pre}
	Feng Du, Shuaishuai Luo, Brenden R. Ortiz, Ye Chen, Weiyin Duan, Dongting Zhang, Xin Lu, Stephen D. Wilson, Yu Song, Huiqiu Yuan, Phys. Rev. B {\bf103}, L220504 (2021).
	
	\bibitem{chengxiaolong}
	Xu Chen, Xinhui Zhan, Xiaojun Wang, Jun Deng, Xiao-bing Liu, Xin Chen, Jian-gang Guo, Xiaolong Chen, Chinese Physics Letters {\bf38}, 057402 (2021).
	
	\bibitem{lishiyan_pre}
	C. C. Zhu, X. F. Yang, W. Xia, Q. W. Yin, L. S. Wang, C. C. Zhao, D. Z. Dai, C. P. Tu, B. Q. Song, Z. C. Tao, Z. J. Tu, C. S. Gong, H. C. Lei, Y. F. Guo, S. Y. Li, arXiv:2104.14487.
	
	\bibitem{yangzhaorong}
	Zhuyi Zhang, Zheng Chen, Ying Zhou, Yifang Yuan, Shuyang Wang, Jing Wang, Hiayang Yang, Chao An, Lili Zhang, Xiangde Zhu, Yonghui Zhou, Xuliang Chen, Jianhui Zhou, Zhaorong Yang, Phys. Rev. B {\bf103}, 224513 (2021).
	
	
	
	\bibitem{luzy}	
	Jian-Feng Zhang, Kai Liu, Zhong-Yi Lu, arXiv:2106.11477.		
	
	

\end{thebibliography}
\end{document}